\documentclass[aps,prd,twocolumn,showpacs,notitlepage,eqsecnum,superscriptaddress,nofootinbib]{revtex4-1}

\usepackage[centertags]{amsmath}
\usepackage{amssymb}
\usepackage{latexsym}
\usepackage{enumerate}
\usepackage{graphicx}
\usepackage{mathrsfs}
\usepackage[colorlinks]{hyperref}
\usepackage{stmaryrd}
\usepackage{import}
\usepackage{tensor}
\usepackage[usenames,dvipsnames]{xcolor}
\usepackage{bm}
\usepackage{multirow}
\usepackage{breakurl}
\usepackage{float}
\usepackage{verbatim}

\allowdisplaybreaks[1]

\definecolor{CiteColor}{rgb}{0,0.5,0}
\hypersetup{citecolor=CiteColor}
\definecolor{RefColor}{rgb}{0.55,0,0}
\hypersetup{linkcolor=RefColor}
\definecolor{darkgreen}{rgb}{0.2,0.7,0.2}

\newcommand{\bi}{\begin{itemize}}
\newcommand{\ei}{\end{itemize}}
\newcommand{\be}{\begin{equation}}
\newcommand{\ee}{\end{equation}}
\newcommand{\nn}{\nonumber}

\renewcommand{\l}{\left(}
\renewcommand{\r}{\right)}
\renewcommand{\a}{\alpha}
\renewcommand{\b}{\beta}
\newcommand{\g}{\gamma}

\renewcommand{\d}{\delta}
\newcommand{\D}{\Delta}
\newcommand{\e}{\epsilon}

\renewcommand{\O}{\Omega}
\renewcommand{\o}{\omega}
\renewcommand{\th}{\theta}

\newcommand{\q}{\quad}

\newcommand{\s}{\sigma}

\newcommand{\ti}{\tilde}

\newcommand{\pa}{\partial}

\newcommand{\bscal}[1]{\boldsymbol{\mathcal{#1}}}

\begin{document}

\title{Highly eccentric inspirals into a black hole}

\author{Thomas Osburn}
\affiliation{Department of Physics and Astronomy, 
University of North Carolina, Chapel Hill, North Carolina 27599, USA}
\author{Niels Warburton}
\affiliation{MIT Kavli Institute for Astrophysics and Space Research, 
Massachusetts Institute of Technology, Cambridge, Massachusetts 02139, USA}
\author{Charles R.~Evans}
\affiliation{Department of Physics and Astronomy, 
University of North Carolina, Chapel Hill, North Carolina 27599, USA}

\begin{abstract}
We model the inspiral of a compact stellar-mass object into a massive nonrotating black 
hole including all dissipative and conservative first-order-in-the-mass-ratio 
effects on the orbital motion.  The techniques we develop allow inspirals with 
initial eccentricities as high as $e\sim0.8$ 
and initial separations as large as $p\sim 50$ 
to be evolved through many 
thousands of orbits up to the onset of the plunge into the black hole.  The 
inspiral is computed using an osculating elements scheme driven by a 
hybridized self-force model, which combines Lorenz-gauge self-force results 
with highly accurate flux data from a Regge-Wheeler-Zerilli code.  The high 
accuracy of our hybrid self-force model allows the orbital phase of the 
inspirals to be tracked to within $\sim0.1$ radians or better.  The difference 
between self-force models and inspirals computed in the radiative 
approximation is quantified.
\end{abstract}


\maketitle

\section{Introduction}
\label{sec:intro}

Relativistic compact binary systems are promising astrophysical sources of gravitational waves. Detection of gravitational waves by ground-based detectors, such as LIGO \cite{LIGO}, VIRGO \cite{VIRGO} or KAGRA \cite{KAGRA}, or future space-based detectors, such as eLISA \cite{Seoane:2013qna}, will be facilitated by accurate theoretical waveform templates. These theoretical templates will also allow the parameters of the source to be determined, which will inform population studies of compact objects as well as allow precision tests of general relativity in the strong-field regime. Producing suitable waveform templates requires solving the two body problem in a general relativistic context which, unlike its Newtonian counterpart, does not have a closed form solution. A number of different techniques exist to approximate solutions to this problem, each applicable to a different class of system depending upon the orbital separation or the mass-ratio of the two bodies.

When the two bodies are widely separated, the post-Newtonian (PN) expansion can be employed \cite{Blan14}. This expansion performs well in the slow adiabatic phase of the inspiral but becomes less accurate as the orbital separation decreases. Once the strong-field regime is entered, for comparable-mass systems, no analytic approximations can be made and the full nonlinear Einstein equations must be numerically solved on a supercomputer \cite{BaumShap10,LehnPret14}. More extreme-mass-ratio systems are beyond the current reach of numerical relativity due to the high resolution requirements around the smaller body and the wide separation of time scales in the problem. In this regime one turns to black hole perturbation theory \cite{Bara09,PoisPounVega11,Thor11}. In addition to the above approaches there is also effective-one-body theory \cite{BuonDamo99,BuonETC09,Damo10}, which incorporates elements from all three of the aforementioned schemes.

In this work we are interested in the inspiral of a stellar-mass compact 
object (such as a black hole, neutron star, or white dwarf) into a 
substantially more massive black hole.  When the binary system consists of a 
supermassive black hole of mass $M\sim 10^5$--$10^7 \, M_\odot$ and a smaller 
compact object of mass $\mu \sim 1$--$10 \, M_\odot$ (so the mass ratio 
$\epsilon=10^{-5}$--$10^{-7}$) the emitted gravitational waves will be in 
the frequency band detectable by space-based detectors such as eLISA.  Such 
extreme-mass-ratio-inspirals (EMRIs) are expected to provide clean tests of 
general relativity in the strong-field regime 
\cite{VigeHugh10,BaraCutl07,BrowETC07,Gair:2012nm} (unspoiled by environmental 
effects \cite{Barausse:2014tra}).  Less extreme mass-ratio binary systems are 
also of interest as they will be observable with Advanced LIGO 
\cite{BrowETC07,AmarETC07}.  For this to occur, intermediate mass black holes 
must exist with masses $M \sim 10^2$--$10^4 M_{\odot}$ \cite{MillColb04}.  
A binary system consisting of an intermediate mass black hole and a smaller 
compact object of mass $\mu \sim 1$--$10 \, M_\odot$ is called an intermediate 
mass-ratio inspiral (IMRI).

Modeling EMRIs and IMRIs is achieved by perturbatively expanding the Einstein field equations in powers of the (small) mass ratio. Typically, the smaller body is modeled as a point particle and the particle's interaction with its metric perturbation gives rise (after regularization) to a self-force that drives the inspiral \cite{MinoSasaTana97,QuinWald97,BaraOri00,DetwWhit03,GralWald08,Poun10}. 
Calculating this self-force has been a major research effort for the past 15 
years that has met with great success, both in computing the gravitational 
self-force 
\cite{BaraSago07,BaraSago10,Akca11,AkcaWarbBara13,DolaBara13,Merlin:2014qda} 
and conservative gauge-invariant quantities 
\cite{Detw08,DolaETC14a,DolaETC14b,Nolan:2015vpa}, which have been compared 
with results from other approaches to the two-body problem 
\cite{BlanETC09,LetiETC11,Akcay:2012ea,Bini:2013rfa,Bini:2014ica,Bini:2014zxa,Isoyama:2014mja,ShahFrieWhit14,Kavanagh:2015lva,ShahPoun15,Akcay:2015pza,VandShah15}.

For computing inspirals it is important to calculate the self-force to high accuracy because in order to detect and accurately extract source parameters from an E/IMRI waveform the phase evolution will need to be tracked to within $\sim 0.1$ radians or less.  This requirement is challenging because from the time an EMRI enters eLISA's passband to when the binary's components merge there is an orbital phase accumulation of order $\e^{-1}\sim 10^5$--$10^7$ radians.  A second 
challenge is the need to calculate the self-force for highly eccentric orbits, 
as we expect astrophysical sources to enter the eLISA passband with 
eccentricities peaked around $e\sim 0.7$--$0.8$ \cite{HopmAlex05}. 

To meet our accuracy goal it is necessary to go beyond leading-order flux 
balance evolutions (so-called radiative or secular approximations) 
\cite{DrasHugh06,FujiHikiTago09} and include conservative and subleading-order 
dissipative corrections to the orbital motion.  This we achieve by using a 
recently developed frequency-domain Lorenz-gauge self-force code 
\cite{OsbuETC14}.  However, the raw output of that code is still not 
sufficient to reach our accuracy goal across the entire parameter space of 
inspirals (especially at high eccentricity).  Instead, as argued in 
\cite{OsbuETC14}, the Lorenz-gauge results can be augmented with high-accuracy 
flux data from a Regge-Wheeler-Zerilli (RWZ) code to produce a hybrid 
self-force scheme.  This paper shows that hybridization in action and confirms 
that our accuracy requirements can be met.  Importantly, by accurately 
reaching eccentricities as high as $e\sim0.8$, the hybrid code breaks a 
barrier where traditionally it was thought that frequency-domain codes 
\cite{WarbBara11,AkcaWarbBara13} must give way to time-domain calculations.

We compute our inspirals by calculating the Lorenz-gauge self-force for over 
9500 geodesics of a Schwarzschild black hole. The hybrid self-force is 
constructed by combining the Lorenz-gauge data with RWZ flux results from over 
40,000 geodesics.  The resulting forces are interpolated across the orbital 
parameter space.  We then evolve our orbits using an osculating element 
scheme.  It is key to point out 
that by using the \emph{geodesic self-force} we are making an approximation.  
The true self-force is a functional of the past history of the 
\emph{inspiraling motion}, whereas in our scheme (and other recent ones 
\cite{WarbETC12,Lackeos:2012de}) we take the self-force at each instance to be that of 
a particle that has moved along a background geodesic for all time.  These 
two self-forces are thought to differ at the first postadiabatic order 
\cite{Pound:2015tma}, and there is ongoing work to quantify the error that 
is induced in the inspiral phase when using this approximation 
\cite{Warb13,Warb14a,Dien15}. As mentioned, the same approach was taken in 
Ref.~\cite{WarbETC12}.

This project is distinguished, however, in several respects. Our inspirals meet 
observationally motivated accuracy goals in contrast with Ref.~\cite{WarbETC12}. 
These accuracy goals are achieved through a novel interpolation scheme, a more 
dense basis of self-force models, parametrization of the orbit in a way that 
accounts for the separatrix, and, as mentioned, a hybridized self-force. We are 
able to cover the full astrophysical range of eccentricities and separations rather 
than the low-eccentricity/small-separation or quasicircular evolutions modeled by 
Refs.~\cite{WarbETC12} and \cite{Lackeos:2012de}, respectively. 
We also introduce a new technique based on Pade approximants to mitigate a 
well-known ill-conditioning problem met when calculating the Lorenz gauge 
self-force \cite{AkcaWarbBara13,OsbuETC14}. Finally, we quantify the phase discrepancy between self-force models and 
inspirals computed in the radiative approximation.

The layout of this paper is as follows.  In Sec.~\ref{sec:effects_of_SF} we 
review how the self-force influences an inspiral and in 
Sec.~\ref{sec:overview} we discuss our approach to computing inspirals using 
the osculating element scheme detailed in Sec.~\ref{sec:osculate}.  In 
Secs.~\ref{sec:lorenz} and \ref{sec:interp} we present our hybridized 
self-force model and its interpolation over the parameter space of geodesics.  
In Sec.~\ref{sec:freq_match} we discuss how to compare inspirals computed 
using the full self-force with inspirals computed using an approximate 
self-force.  Our main results are then presented in Sec.~\ref{sec:res} and we 
conclude with some final remarks in Sec.~\ref{sec:con}.  Throughout this paper 
we set $c = G = 1$, use metric signature $(-+++)$ and the sign conventions of 
Misner, Thorne, and Wheeler \cite{MisnThorWhee73}.

\section{Effects of the self-force on an inspiral}
\label{sec:effects_of_SF}

The smaller body's interaction with its own metric perturbation gives rise to 
a self-force that causes the motion to deviate from a geodesic of the 
background spacetime of the larger black hole.  In practice we calculate 
this self-force perturbatively, expanding the Einstein field equations in 
powers of the mass ratio, $\epsilon = \mu/M \ll 1$.  How we compute the 
self-force from a suitably regularized metric perturbation will be discussed 
in Sec.~\ref{sec:lorenz}.  The self-force drives the motion off of a 
background geodesic\footnote{Alternatively, the motion can be 
considered as a geodesic in a regular effective space-time 
\cite{DetwWhit03,PoisPounVega11}.  Here we use the ``forced motion in the 
background spacetime'' picture but both viewpoints are equally valid.} in the 
following way
\begin{align}
\label{eqn:forced}
\mu u^\beta \nabla_\beta u^\alpha 
= F^\alpha_{(1)} + F^\alpha_{(2)} + \mathcal{O}(\epsilon^4) ,
\end{align}
where $u^\a$ is the body's four-velocity and $\nabla$ denotes the covariant 
derivative with respect to the background metric.  By $F^\alpha_{(n)}$ we 
denote the $n$th-order self-force, i.e., the part 
proportional to the $n+1$ power of the mass ratio.  Alternatively, we may use
the covariant form of \eqref{eqn:forced} for the evolution of $u_{\alpha}$, 
which requires the covariant form of the self-force $F_{\alpha}$.  This latter 
form of the equation of motion, it turns out, plays an important role in our 
hybrid method, as described in Sec.~\ref{sec:lorenz}.

\begin{figure}
\includegraphics[width=8.5cm]{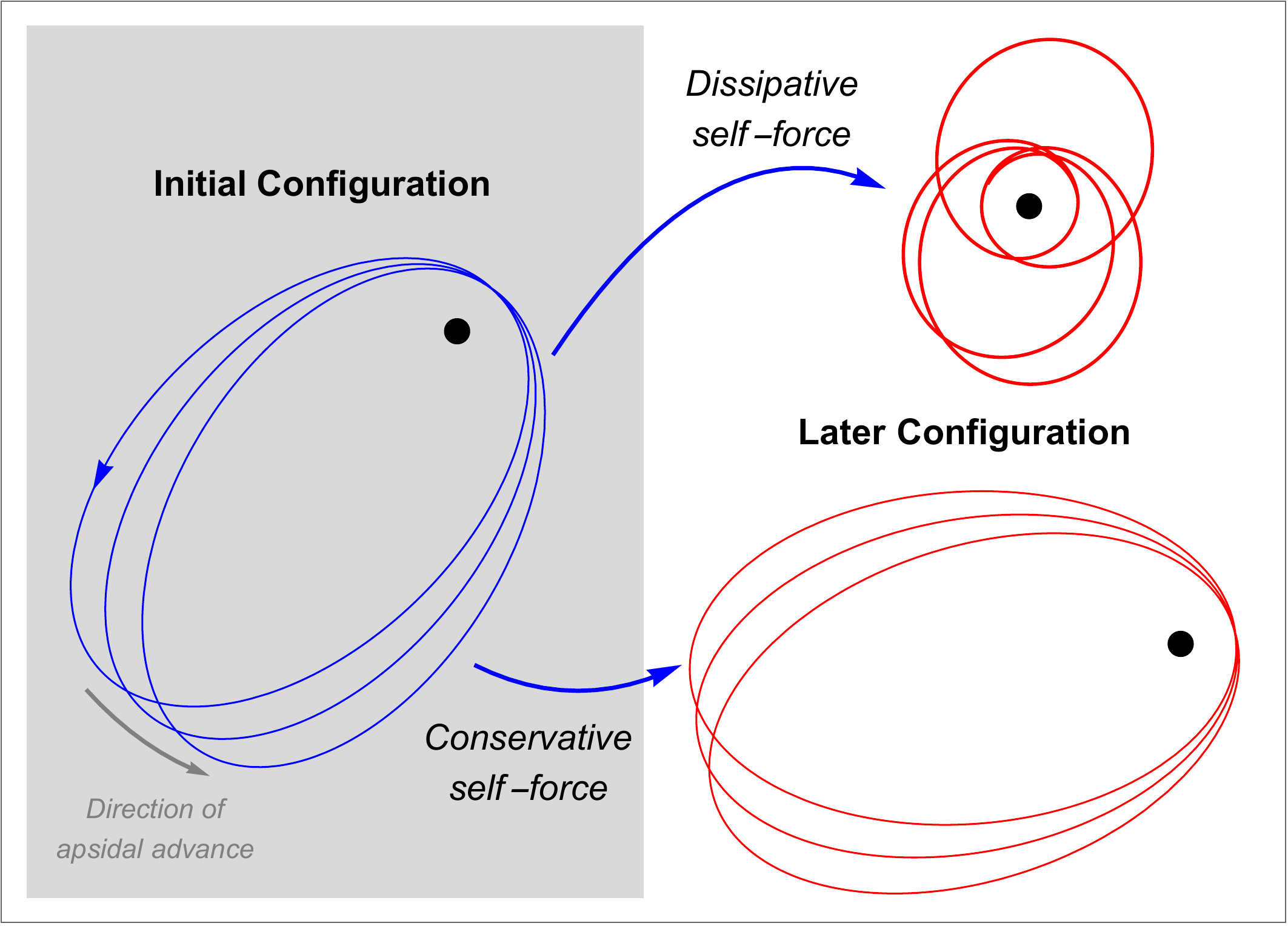}
\caption{Illustration of dissipative and conservative self-force influences 
on the inspiral.  On the left, the motion of the compact body around the 
central black hole is taken to be counterclockwise, as is then the apsidal 
advance of the orbit.  On the right top, the dissipative self-force is 
responsible for the secular decay of energy and angular momentum, which 
causes the orbit to shrink and shed eccentricity.  In contrast (right bottom), 
the conservative self-force does not affect the long-term average of the 
orbital constants.  Instead it is responsible for a slower than usual apsidal 
advance \cite{BaraSago11} and tiny periodic changes in the shape of the 
eccentric orbit.  Both effects occur simultaneously during a physical 
inspiral.}\label{fig:consdiss}
\end{figure}

In the geodesic self-force approximation we can in addition split the force, 
at each order, into a conservative part, $F^\a_\text{cons}$, attributed to 
the time-symmetric part of the gravitational field and a dissipative part, 
$F^\a_\text{diss}$, due to the time-antisymmetric part of the gravitational 
field
\begin{align}
F^\a = F^\a_\text{diss} + F^\a_\text{cons} .
\end{align}
The dissipative part is responsible for radiation reaction effects such as 
the decay of orbital energy and angular momentum.  The conservative part 
perturbs the orbital parameters, but does not cause a secular decay of the 
orbit.  See Fig.~\ref{fig:consdiss} for an illustration of these effects. 

The dissipative self-force can be further split into two parts: an adiabatic 
part $F^{\alpha\text{(diss)}}_\text{ad}$, whose components vary slowly over an 
inspiral on the radiation reaction time scale and represents some average over 
the orbital time scale, and an oscillating part 
$F^{\alpha\text{(diss)}}_\text{osc}$, whose components oscillate on the 
orbital time scale.  We can thus write for the full self-force
\begin{align}
\label{eq:F_ad_and_osc}
F^\alpha &= 	F^{\alpha\text{(diss)}}_\text{ad} + F^\alpha_\text{osc} , \\
F^\alpha_\text{osc} &\equiv 
F^{\alpha\text{(diss)}}_\text{osc} + F^{\alpha\text{(cons)}} .
\end{align}
Unfortunately, the adiabatic/oscillatory split is ambiguous at this point.  
The general intent is to take
\begin{align}
\label{eqn:splitFad}
F^{\alpha\text{(diss)}}_\text{ad} &\equiv
\langle F^{\alpha\text{(diss)}} \rangle , \\
F^{\alpha\text{(diss)}}_\text{osc} &=
F^{\alpha\text{(diss)}} -
\langle F^{\alpha\text{(diss)}} \rangle , 
\end{align}
but to be precise this requires a specific definition for the average 
$\langle \quad \rangle$ over the orbital time scale.  The ambiguity comes 
because the averaging can be performed with respect to different curve parameters 
and, again because of the orbital eccentricity, there is a difference in 
averaging contravariant versus covariant components.  See the discussion by 
Pound and Poisson in \cite{PounPois08b} on this ambiguity and its effect on 
defining an ``adiabatic,'' ``secular,'' or ``radiative'' approximation.  In 
this paper, even though we avoid making the adiabatic approximation, we 
nonetheless have a use for this decomposition in defining our hybrid scheme, 
and single out a particular definition for the averaging process.  This 
specific choice is discussed further below and in Sec.~\ref{sec:lorenz}.  

Assuming that some definition is adopted, at any moment in time the adiabatic 
and oscillatory parts will be comparable in size.  However, if we compute the 
oscillatory part along a bound geodesic of the background spacetime (i.e., 
compute the self-force but do not actually apply it), the average over one 
orbit of $F^{\alpha\text{(diss)}}_\text{osc}$ vanishes by construction (this 
is true also of $F^{\alpha\text{(cons)}}$).  If instead the self-force is 
applied and the orbit allowed to evolve, then 
$F^{\alpha\text{(diss)}}_\text{osc}$ (and $F^{\alpha\text{(cons)}}$) will 
nearly average to zero over one radial orbital period, with the residual 
being of order $\mathcal{O}(\epsilon)$ relative to a typical instantaneous 
magnitude.  

The smallness of this average implies a gradual, adiabatic inspiral, and is a 
needed justification for using the geodesic self-force.  A number of authors 
have considered how these different parts of the self-force influence the 
inspiral phase \cite{HughETC05,DrasFlanHugh05,Tanaka:2006} with one of the 
most rigorous discussions given by Hinderer and Flanagan \cite{HindFlan08}.  
We review several key results and highlight where previous work has employed 
the various components of the self-force in computing inspirals.

With an E/IMRI there is a large accumulation of orbital phase from the point 
when the binary enters, say, the eLISA passband until merger.  The 
leading-order part to the orbital phase enters at $\mathcal{O}(\epsilon^{-1})$ 
and is driven by the abovementioned adiabatic, first-order-in-the-mass-ratio, 
dissipative self-force $F^{\alpha\text{(diss)}}_{(1)\text{ad}}$.  
Conveniently, this component of the self-force can be related to the 
orbit-averaged asymptotic fluxes\footnote{Flux balance arguments allow the 
evolution of the orbital energy and angular momentum to be computed.  For 
generic orbits in Kerr spacetime the evolution of the Carter constant is 
computed using methods introduced by Mino \cite{Mino03,SagoETC05,SagoETC06}}, 
which sidesteps the need for a more complicated, local calculation of the 
self-force from the metric perturbation at the particle.  A number of authors 
have used this approach to calculate the leading-order phase evolution of 
generic inspirals into Kerr black holes \cite{DrasHugh06,FujiHikiTago09}, 
though at the cost of missing some effects available within the first order 
perturbation.

In a regular perturbation calculation, the next effects in the cumulative 
phase would be at $\mathcal{O}(\epsilon^0)$.  However, it is known that for 
generic inspirals in Kerr spacetime, certain resonant configurations will 
occur that contribute to the cumulative orbital phase at 
$\mathcal{O}(\epsilon^{-1/2})$.  These transient resonances take place when 
the radial and polar orbital frequencies are in a low-integer ratio 
\cite{FlanHind10} and will generically occur a few times during any inspiral 
\cite{Ruangsri:2013hra}.  Resonant orbits are an active area of research 
\cite{Brink:2013nna,vandeMeent:2013sza,Brink:2015roa} but will not be 
considered further in this work as we concentrate on inspirals in 
Schwarzschild spacetime.

The next contributions to the orbital phase lie at $\mathcal{O}(\epsilon^0)$.  
These include the conservative part of the first-order self-force, the 
oscillatory part of the dissipative first-order self-force, and the adiabatic 
part of the dissipative second-order self-force.  (At this order, there is 
also expected to be a difference between using the geodesic self-force 
instead of the true self-force.)  The first two contributions require a local 
calculation of the self-force and in recent years there has been great 
progress evaluating these quantities 
\cite{BaraSago07,BaraSago10,Akca11,AkcaWarbBara13,OsbuETC14,Merlin:2014qda}.  
The first low-eccentricity inspirals in Schwarzschild spacetime computed 
incorporating these two components were presented in Ref.~\cite{WarbETC12}.  
The evolution of quasicircular inspirals has also been explored 
\cite{Lackeos:2012de}.  As yet there have been no calculations of the 
second-order-in-the-mass-ratio self-force but the appropriate formalism and 
calculation techniques are emerging 
\cite{Poun12a,Gral12,Detw12,PounMill14,WarbWard14,WardWarb15}.

To summarize, the influence of each component of the self-force on the 
phase of the waveform in the inspiral, as measured, for example, by using the 
cumulative radial phase $\Phi_r$ as a proxy, is 
\begin{align}
\label{eq:phase_scaling}
\Phi_r = &\underbrace{\kappa_{0} \; 
\epsilon^{-1}}_\text{adiabatic: $F^{\alpha\text{(diss)}}_{(1)\text{ad}}$} 
+\underbrace{\kappa_{1/2} \; 
\epsilon^{-1/2}}_\text{resonances (Kerr only)} \\ \nn
&+ \underbrace{\kappa_{1} \; 
\epsilon^0}_{\substack{\text{post-1-adiabatic:} \\ 
\text{$F^{\alpha\text{(cons)}}_{(1)} 
+ F^{\alpha\text{(diss)}}_{(1)\text{osc}} 
+ F^{\alpha\text{(diss)}}_{(2)\text{ad}}$} }} + \cdots ,
\end{align}
where the $\kappa$ coefficients are dimensionless, of order unity, and depend
on the ingress and egress (or merger) frequencies in a particular detector, 
but not on the mass ratio $\epsilon$.  The adiabatic dissipative part of the 
self-force comes in at lower order than 
the remaining parts of the self-force, and accordingly must be computed with 
greater accuracy in order to affect the phase error at the same level.  Even 
though our present calculations account for all first-order-in-the-mass-ratio 
contributions in the geodesic self-force, we purposefully make the split into 
adiabatic dissipative and oscillatory dissipative parts so that these two 
pieces can be computed, in the hybrid scheme, to their separate fractional 
accuracies.

\section{Overview of our approach}\label{sec:overview}

Formally, the self-force is a functional of the entire past history of the 
inspiral.  Letting $z^\alpha(\tau)$ denote the particle's inspiraling 
worldline, with $\tau$ being proper time, we can write the self-force as 
$F^\alpha(\tau) \equiv F^\alpha[z^\alpha(\tau' < \tau)]$.  In order to 
compute an inspiral in a self-consistent manner one must solve for the 
worldline using Eq.~\eqref{eqn:forced} whilst simultaneously computing the 
perturbation in the gravitational field and its effects in generating the 
local self-force.

To date, such a self-consistent inspiral has only been computed for the case 
of a scalar particle \cite{DienETC12}.  Instabilities with the time-domain 
evolution of the low $l$-modes of the Lorenz-gauge self-force currently stand 
in the way of computing self-consistent inspirals in the gravitational case 
(see Ref.~\cite{DolaBara13} for a discussion of these gauge instabilities).  
This provides part of the motivation for using the geodesic self-force 
approach to computing the inspiral.  A secondary motivation comes from noting 
the high computational cost of evolving inspirals in the time-domain.  
Currently available technology (for the case of a scalar particle) allows for 
the computation of inspirals with a few tens of periastron passages at the 
cost of weeks of runtime on hundreds of CPU cores \cite{DienPrivComm}.  Certainly, in the near 
future, such time-domain approaches will not be extensible to computing the 
many hundreds of thousands of periastron passages that occur in an 
astrophysical EMRI.  Furthermore, it will be required to compute many 
thousands of inspirals in order to construct a suitably dense template bank 
of waveforms for use in matched filtering searches.  In the method we employ 
here, a single preprocessing step takes a few thousand CPU hours and once 
that is complete each inspiral can be computed in a matter of minutes. 

The geodesic self-force approach stems from the key observation that, as an 
EMRI evolves adiabatically, the inspiral is closely approximated at each 
moment by a background geodesic that is tangent to the true (inspiralling) worldline.  At 
each moment the true self-force is approximated by the (geodesic) self-force 
that would exist if the motion were not driven off the background geodesic.  
Differences between the true inspiral and the background geodesic are greatest 
in the distant past, and the tail integral that gives rise to the local 
self-force is expected to have falling contributions for $\tau' \ll \tau$.  
Similar higher-order effects due to differences between true evolution and 
fixed-orbit calculations occur in post-Newtonian theory \cite{ArunETC08a}.  In 
this picture, inspirals are evolved by replacing $F^\alpha$ in 
Eq.~\eqref{eqn:forced} with 
$F_G^\alpha(\tau) \equiv F^\alpha [z^\alpha(\tau);z_G^\alpha(\tau')]$ where 
$z_G^\alpha(\tau')$ is the worldline of the background geodesic tangent to 
$z(\tau)$.  Working with the geodesic self-force has a key advantage that 
during the inspiral phase the tangent geodesic is bound and strictly 
periodic.  The periodic nature of the tangent geodesic allows for an efficient 
frequency-domain calculation of the self-force \cite{AkcaWarbBara13,OsbuETC14}. 
Moreover, working in the frequency-domain avoids the gauge instabilities 
observed in time-domain evolutions.  Although frequency-domain codes can 
compute self-force data rapidly, they are not sufficiently quick to allow 
direct on-the-fly inspiral evolutions.  Instead we interpolate the self-force 
data over the applicable range of the orbital parameter space.  A new and 
efficient interpolation procedure is presented in Sec.~\ref{sec:interp}.

Previous applications of geodesic self-force evolution 
\cite{WarbETC12,Lackeos:2012de} only probed small eccentricities 
($e \lesssim 0.2$).  As astrophysical EMRIs are expected to have high 
eccentricities \cite{HopmAlex05}, we have worked to expand the range of the 
technique to model eccentricities up to $e \lesssim 0.8$.  Furthermore, 
a self-force model must be sufficiently accurate to capture correctly the 
phase evolution of the inspiral for matched-filtering purposes.  As the 
previous section noted in Eq.~\eqref{eq:phase_scaling}, we do not need to 
know all pieces of the self-force to the same accuracy.  (This is fortunate 
since some parts of the self-force are more challenging to compute than 
others.)  This motivates the hybrid self-force method discussed in 
\cite{OsbuETC14}.  As summarized in Table~\ref{table:required_accuracies}, 
the most sensitive part of the self-force--the adiabatic dissipative part--can 
be calculated from fluxes obtained with a very accurate RWZ code, while the 
oscillatory part of the dissipative self-force and the conservative part can 
be computed with a Lorenz gauge code \cite{OsbuETC14}.  The separate required 
accuracies are listed in Table~\ref{table:required_accuraries}.  How data 
from the two codes are combined is discussed in Sec.~\ref{sec:hyb}.

\begin{table}
  \caption{The required accuracies for an archetypal EMRI system with a 
massive black hole of mass $10^6 M_\odot$ orbited by a stellar mass black 
hole of mass $10 M_\odot$ ($\epsilon = 10^{-5}$).  The scaling of the phase 
evolution from Eq.~\eqref{eq:phase_scaling} implies the accuracy with which 
we need to obtain the self-force.  Row two of the table shows the precision 
in the self-force required to track the phase evolution to within 
$\sim 0.1$ radians.  Row three gives the codes we use to compute the various 
components, and row four shows the precision in the output data from these 
codes.  The wide range in precision of the Lorenz-gauge code is a function 
of the orbital eccentricity.  At present there are no codes able to compute 
the second-order self-force in the strong-field (though 
Ref.~\cite{Burko:2013cca} uses a PN flux formula to explore the effects of 
the second-order self-force upon an quasicircular evolution).}
\label{table:required_accuracies}\label{table:required_accuraries}
\begin{tabular}{ l | c | c | c }
\hline
\hline
	& $F^{\alpha\text{(diss)}}_{(1)\text{ad}}$	& $F^{\alpha}_{(1)\text{osc}}$		&	$F^{\alpha\text{(diss)}}_{(2)\text{ad}}$ \\
    \hline
    Required accuracy 	& $10^{-7}$ 								& $10^{-2}$							& $10^{-2}$ 		\\ 
    Code 				& RWZ \cite{HoppEvan10,HoppETC15}						& Lorenz \cite{OsbuETC14}			& $\cdots$		\\
    Code accuracy 		& $10^{-10}$-$10^{-9}$ 						& $10^{-7}$-$10^{-3}$ 				& $\cdots$		\\
    \hline
  \end{tabular}
\end{table}

Finally, we must address how the geodesic self-force approximation will 
influence the phasing of the modeled orbit.  As mentioned above, arguments 
have been made that the error will enter at $\mathcal{O}(\epsilon^0)$ 
(see Sec.~1.5.6 of Ref.~\cite{Pound:2015tma}).  This might seem discouraging 
as the geodesic self-force approximation is introducing an error in the phase 
at the same order as the oscillatory conservative and dissipative effects we 
have worked hard to include.  The only way to assess how problematic this is 
to our approach is to compare our evolution with a fully self-consistent 
one.  As mentioned, this is not yet possible for the gravitational case.  
Work is ongoing, however, to make this comparison for scalar self-force 
evolutions.  Preliminary work comparing the self-consistent time-domain 
code of Ref.~\cite{DienETC12} and a geodesic scalar self-force inspiral code 
constructed using the techniques of Ref.~\cite{WarbBara11} indicates that, 
although the phase error might enter at $\mathcal{O}(\epsilon^0)$, the 
coefficient must be small (in fact so small it has yet to be measured despite 
concerted effort \cite{Warb13,Warb14a,Dien15}).

\section{Osculating element description of motion}
\label{sec:osculate}

Our approach is to solve Eq.~\eqref{eqn:forced} using the geodesic self-force 
as the forcing term.  Similar to Newtonian celestial mechanics calculations, 
we recast the equation of motion into one for the evolution of osculating 
elements of the inspiral.  The resulting inspiral can be immediately 
interpreted in a geometric manner and the numerical output from the self-force 
codes can be more easily linked to the long-term evolution code.

In the osculating element approach the true (accelerated) worldline, 
$z(\tau)$, is taken to be tangent to a background geodesic worldline 
$z_G(\tau)$ at each time $\tau$.  As the true worldline advances, the 
parameters of the background geodesic smoothly evolve.  At each instance the 
tangent (or ``osculating'') geodesic is characterized \cite{PounPois08b} by a 
set of orbital elements $I^A$, with the true worldline represented by a 
continuous sequence of elements $I^A(\tau)$.  With the four-velocity of the 
tangent geodesic given by $u^\a_G(I^A,\tau) = \pa_\tau z^\a_G(I^A,\tau)$, we 
can write
\begin{align}
	z^\a(\tau) = z^\a_G(I^A,\tau), \q\q u^\a(\tau) = u^\a_G(I^A,\tau) .
\end{align}
We thus seek equations of motion for the set of osculating elements.  This 
procedure was first outlined by Pound and Poisson \cite{PounPois08b} for 
motion about a Schwarzschild black hole.  Extension of the idea to motion in 
Kerr spacetime was given by Gair \emph{et al.}~\cite{GairETC11}.  The resulting 
equations of motion take the form
\begin{align}
	\label{eqn:Ievolve}
	\frac{\pa z^\a_G}{\pa I^A} \frac{\pa I^A}{\pa\tau} = 0 , \q\q\q 
	\mu \frac{\pa u^\a_G}{\pa I^A}\frac{\pa I^A}{\pa\tau} = F^\a .
\end{align}
Our explicit choices for $I^A$ for bound motion and the resulting equations 
of motion are given in the next subsection.  It is important to note that 
the osculating element approach is simply a recasting of 
Eq.~\eqref{eqn:forced} and is valid for any forcing 
term\footnote{so long as the tangent geodesic remains bounded}; no small 
forcing approximation is made. 

\subsection{Bound geodesics in Schwarzschild spacetime}
\label{sec:geo}

We consider in this paper bound and eccentric motion around a Schwarzschild 
black hole.  Schwarzschild coordinates $x^{\a} = (t,r,\theta, \varphi )$ are 
adopted, in which the line element takes the standard form
\begin{align}
ds^2 = -f \, dt^2 + f^{-1} dr^2 + r^2 
\left( d\theta^2 + \sin^2\theta \, d\varphi^2 \right) ,
\end{align}
where $f(r) = 1 - 2M/r$.  The geodesic worldline is given by a set of functions
$z_G^{\a}(\tau) 
=\left[t_p(\tau),r_p(\tau),\th_p(\tau),\varphi_p(\tau)\right]$, parametrized 
by (for example) proper time $\tau$.  Without loss of generality the motion 
is confined to the equatorial plane, $\theta=\pi/2$.  The geodesic 
four-velocity $u^\a_G$ is given by
\begin{align}
\label{eqn:fourvelocity}
	u_G^\a = \l \frac{{\cal{E}}}{f_p}, u^r_G, 0, \frac{{\cal{L}}}{r_p^2} \r,
\end{align}
where $f_p\equiv f(r_p)$ and ${\cal{E}}$ and ${\cal{L}}$ are the specific energy and angular 
momentum, respectively.  The constraint $u_G^\a u^G_\a = -1$ yields an 
expression for $u^r_G$:
\begin{align}
\label{eqn:ur}
\l u^{r}_G \r^2 = \mathcal{E}^2-f_p\left(1+\frac{\mathcal{L}^2}{r_p^2}\right) .
\end{align}
We parametrize the geodesic with the eccentricity, $e$, and semilatus 
rectum, $p$, which are related to the radial turning points $r_{\rm min}$ 
and $r_{\rm max}$ via
\begin{align}
	\label{eqn:defeandp}
	p = \frac{2 r_{\rm max} r_{\rm min}}{M (r_{\rm max} + r_{\rm min})} , 
	\q \q
	e = \frac{r_{\rm max} - r_{\rm min}}{r_{\rm max} + r_{\rm min}}.
\end{align}
Equation~\eqref{eqn:defeandp} and the roots of Eq.~\eqref{eqn:ur} give the 
relationship between ($p$,$e$) and ($\mathcal{E}$,$\mathcal{L}$):
\begin{align}
\label{eqn:specEL}
\mathcal{E} &= \sqrt{\frac{(p-2)^2-4 e^2}{p(p-3-e^2)}}, \q\q\;\;
\mathcal{L} = \frac{pM}{\sqrt{p-3-e^2}} .
\end{align}
Orbits are bound when $e<1$ and are stable when $p>6+2e$. 

In self-force calculations it is convenient to reparametrize the orbital 
motion (i.e., all the curve functions) with the relativistic anomaly $\chi$ 
\cite{Darw59}, defined so that
\begin{align}\label{eq:r_chi}
r_p(\chi) &= \frac{p M}{1+e \cos\left[\chi-\chi_0\right]} .
\end{align}
The parameter $\chi_0$ specifies the value of $\chi$ at pericentric passage.  

Equation~\eqref{eq:r_chi} can be used with Eqs.~\eqref{eqn:fourvelocity} and 
\eqref{eqn:specEL} to derive the following initial value equations for the 
development of the orbit 
\begin{align}
	\label{eqn:tau}
	\frac{d\tau_p}{d\chi} &= \frac{Mp^{3/2}}{(1+e \cos v)^2}\sqrt{\frac{p-3-e^2}{p-6-2 e \cos v}} ,\\
	\label{eqn:t}
	\frac{dt_p}{d\chi} &= \frac{r_p^2}{M(p-2-2 e\cos v)}\sqrt{\frac{(p-2)^2-4e^2}{p-6-2 e\cos v}} , \\
	\label{eqn:phi}
	\frac{d\varphi_p}{d\chi} &= \sqrt{\frac{p}{p-6-2 e\cos v}} ,
\end{align}
where $v\equiv \chi-\chi_0$.  Without loss of generality we can choose 
initial conditions $\varphi_p|_{\chi=0} = 0$, $t_p|_{\chi=0} = 0$, 
$\tau_p|_{\chi=0} = 0$, in which case changes in $\chi_0$ serve, for example, 
to alter the orientation of the orbit.

The periods of one radial libration measured in $t$ and $\tau$ are denoted 
by $T_r$ and $\mathcal{T}_r$, respectively.  They are given by
\begin{align}
	T_r = \int_0^{2\pi} \frac{dt_p}{d\chi} d\chi, \q\q\q	\mathcal{T}_r = \int_0^{2\pi} \frac{d\tau_p}{d\chi} d\chi .
\end{align}
The amount of azimuthal angle accumulated in one radial period, $T_r$, is 
given by
\begin{align}
	\D \varphi = \int_0^{2\pi} \frac{d\varphi_p}{d\chi}d\chi .
\end{align}
Each orbit has associated with it two fundamental frequencies.  One is a libration-type 
frequency associated with the radial motion and the other is a rotation-type 
frequency associated with the average rate at which the orbital azimuthal 
angle accumulates.  These two frequencies are defined via
\begin{align}\label{eq:freqs}
\O_r \equiv \frac{2\pi}{T_r}, \q\q\q\q \O_\varphi \equiv \frac{\D\varphi}{T_r} .
\end{align}

\subsection{Evolution of orbital elements}

For the osculating element scheme, the set of orbital elements we evolve 
are 
\begin{align}
	I^A = (p,e,\chi_0,t_p,\varphi_p) .
\end{align}
The elements $(p,e)$ are ``principal elements'' that describe the spatial 
shape of the tangent geodesic but not its orientation.  The orientation of 
the orbit is set by the ``positional element'' $\chi_0$.  The last two 
elements $(t_p,\varphi_p)$ track the evolution of the time and angular 
coordinate of the orbit.

The evolution of $I^A$ follows from Eqs.~\eqref{eqn:Ievolve}, 
\eqref{eqn:tau}, \eqref{eqn:t}, and \eqref{eqn:phi}.  We also use the 
orthogonality of the self-force and the four-velocity, 
$F^\alpha u_\alpha = 0$, to manipulate how the components of $F^\a$ appear 
in the equations.  The following is our formulation of the evolution 
equations for $e$, $p$, and $\chi_0$
\begin{align}
	\label{eqn:dedchi}
	\frac{de}{d\chi} &= \frac{a_{(t)}(e,p,v) F^t+a_{(\varphi)}(e,p,v) F^{\varphi}}{\mu\, q(e,p,v)} , \\
	\label{eqn:dpdchi}
	\frac{dp}{d\chi} &= \frac{b_{(t)}(e,p,v) F^t+b_{(\varphi)}(e,p,v) F^{\varphi}}{\mu\, q(e,p,v)} , \\
	\label{eqn:dchi0dchi}
	\frac{d\chi_0}{d\chi} &= \frac{c_{(r)}(e,p,v) F^r+c_{(\varphi)}(e,p,v) F^{\varphi}}{\mu\, q(e,p,v)} ,\\
	a_{(t)} &\equiv Mp(p-3-e^2)(6+2e^2-p)(1+e\cos v)^2 \notag
	\\&\q\q\q \times(2-p+2e\cos v)\sqrt{(p-2)^2-4e^2},
	\\
	a_{(\varphi)} &\equiv M^2 p^{5/2}(1-e^2)(3+e^2-p) \notag
	\\& \q\q\q\q\q\q \times\big[4e^2+(p-6)(p-2)\big] ,
	\\
	b_{(t)} &\equiv 2M\,e\,p^2(3+e^2-p)(p-2-2e\cos v) \notag
	\\&\q\q\q\q \times(1+e\cos v)^2\sqrt{(p-2)^2-4e^2},
	\\
	b_{(\varphi)} &\equiv 2M^2e\,p^{7/2}(p-4)^2(p-3-e^2),
	\\
	c_{(r)} &\equiv Mp^2(3+e^2-p)(2e+(p-6)\cos v) \notag
	\\&\q\q\q \times(1+e\cos v)^2\sqrt{p-6-2e\cos v},
	\\
	c_{(\varphi)} &\equiv M^2p^{5/2}\sin v (3+e^2-p)\Big(2(p-6)(3-p) \notag
	\\&\q\;+e\cos v\big[(4e^2-(p-6)^2)+2e(p-6)\cos v\big]\Big),
	\\
	q &\equiv e(p-6-2e)(p-6+2e)\notag
	\\&\q\q\q\times(1+e\cos v)^4\sqrt{p-6-2e\cos v}.
\end{align}
The equations for $t_p$ and $\varphi_p$ are unchanged from 
Eqs.~\eqref{eqn:t} and \eqref{eqn:phi}.  See Ref.~\cite{PounPois08b} for the 
detailed derivation of an equivalent set of evolution equations for the 
osculating elements.

Specifying the initial values of the elements $I^A$ is equivalent to 
specifying the initial position and velocity on Eq.~\eqref{eqn:forced}.  For 
motion in a plane there are three initial positions and two initial velocities 
(three minus the one for the normalization condition 
$u^\alpha_G u_\alpha^G = -1$), which matches the number of initial values 
we specify for $I^A$.

For a long-term evolution, all of the differential equations 
\eqref{eqn:dedchi}, \eqref{eqn:dpdchi}, \eqref{eqn:dchi0dchi}, 
\eqref{eqn:t}, and \eqref{eqn:phi} are integrated simultaneously [along 
with \eqref{eqn:tau} if desired], while continually updating the self-force 
components as derived from the instantaneous background geodesic.  

\section{Calculation of the Lorenz gauge self-force with a hybrid scheme}
\label{sec:lorenz}

Recent frequency-domain codes are now able to rapidly compute the self-force 
in Lorenz gauge along a geodesic orbit \cite{AkcaWarbBara13,OsbuETC14}.  In 
this work we use the code presented in Ref.~\cite{OsbuETC14} but with an 
improvement to the source integration method described in 
Ref.~\cite{HoppETC15}.  Unfortunately, for some high eccentricity inspirals 
that interest us here, that code is still not sufficiently accurate to reach 
the accuracy requirements laid out in Table \ref{table:required_accuraries}.  
However, the amount by which each part of the self-force influences the inspiral 
phasing suggests a solution: use a highly accurate RWZ code to compute 
gauge-invariant fluxes and then use that data to obtain the leading-order 
(orbit-averaged) contribution, $F^{\alpha\text{diss}}_{(1)\text{ad}}$, of the 
self-force.  The remaining parts of the first-order self-force are then 
supplied by the Lorenz-gauge code.  This ``hybrid self-force'' approach 
was sketched out in Sec.~V.B of Ref.~\cite{OsbuETC14}.  We give further 
details below in Sec.~\ref{sec:hyb}.  First, however, we briefly outline 
the Lorenz-gauge self-force and describe an added improvement that has been 
made to the Lorenz-gauge code that helped increase its range of applicability 
and reduced its runtime.

\subsection{Lorenz-gauge self-force}

\label{sec:mp}

The finite mass of the small body induces a perturbation over the background 
metric $g_{\mu \nu}$.  Working to first-order-in-the-mass-ratio, we may write 
the full spacetime metric as ${\rm g}_{\mu\nu} = g_{\mu\nu} + h_{\mu\nu}$, 
with metric perturbation (MP) $h_{\mu\nu}$.  Defining the trace-reversed 
MP by ${\Bar h}_{\mu\nu} \equiv h_{\mu\nu} - \tfrac{1}{2} g_{\mu\nu} h$ 
(with $h = h_{\alpha\beta} \, g^{\alpha\beta}$), the Lorenz-gauge condition 
is given by
\begin{align}
	\label{eqn:tensor_gauge}
	\nabla_\nu\tensor{\Bar h}{^\mu^\nu} = 0 ,
\end{align}
where $\nabla$ is compatible with the background metric.  The Lorenz-gauge 
linearized Einstein equations for the first-order-in-the-mass-ratio MP is 
then given by
\begin{align}
	\Box \bar{h}_{\mu\nu} + 2 \tensor{R}{^\alpha_\mu^\beta_\nu} \, \bar{h}_{\a\b} 
	= -16\pi T_{\mu\nu},
	\label{eqn:LGeqns}
\end{align}
where $\Box \equiv g^{\a\b} \nabla_{\a} \nabla_{\b}$, $R^\alpha_{\,\;\mu\beta\nu}$ 
is the Riemann tensor in the background and $T_{\mu\nu}$ is the stress-energy 
tensor.  The last is taken to be that of a point-mass moving along a fixed, 
bound geodesic of the background spacetime.

The spherical symmetry of the background Schwarzschild spacetime and the 
periodicity of the source allow for a tensor spherical-harmonic and Fourier 
decomposition of Eq.~\eqref{eqn:LGeqns} that fully decouples the individual 
tensor-harmonic and Fourier modes (though for each mode the metric 
perturbation amplitudes generally remain coupled).  The field equations for 
each mode are reduced to a coupled set of ordinary differential equations 
(ODEs), which we solve numerically with suitable boundary conditions to 
construct the retarded solution.  Within this procedure each mode of the 
retarded MP is finite at the particle's location, but the sum over modes 
diverges there.  We employ the method of extended homogeneous solutions to 
ensure the Fourier sum converges exponentially \cite{BaraOriSago08} and 
construct the self-force using the mode-sum regularization scheme of Barack 
and Ori \cite{BaraOri00}.  We also employ additional regularization parameters 
from Ref.~\cite{HeffOtteWard12a} that speed up the convergence of the 
mode-sum.  Full details of the code we use are given in 
Refs.~\cite{OsbuETC14,HoppETC15}.

One challenge with the frequency-domain Lorenz gauge method arises when 
constructing the inhomogeneous solutions from a suitable basis of homogeneous 
solutions using the standard variation of parameters approach.  In this 
approach, a Wronskian matrix of homogenenous solutions is assembled, which 
must be inverted.  This matrix becomes ill-conditioned when the mode 
frequency $\o=m\,\O_\varphi+n\,\O_r$ is small, where $m$ and $n$ are integers 
\cite{AkcaWarbBara13}.  Arbitrarily small frequencies are encountered in the 
neighborhood of orbital resonances where the ratio $\O_\varphi/\O_r$ is a 
rational number.  Without any algorithm to alleviate this issue the 
smallest-frequency modes that can be computed with machine precision are 
typically around $|\o M|\gtrsim 10^{-3}$.  Akcay \emph{et al.}~employed novel 
techniques to handle frequencies as small as 
$|\o M|\gtrsim 10^{-4}$ \cite{AkcaWarbBara13}.  
Even with this improved limit, large fractions of orbital parameter space 
remain excluded from accurate calculation.  Osburn \emph{et al.}~\cite{OsbuETC14} 
used quad-precision numerical integration and other novel techniques to 
handle frequencies as low as $|\o M|\gtrsim 10^{-6}$ \cite{OsbuETC14}.  
Unfortunately, that improvement comes at considerable added computational 
expense. 

As an alternative, we developed a new method to calculate asymptotic boundary 
conditions that utilizes the diagonal Pade approximant (DPA).  With this 
change, we are able to handle frequencies as small as $|\o M|\gtrsim 10^{-5}$ 
while avoiding use of quad-precision numerical integration.  We outline the 
method here as it is applied in the odd-parity sector, though the same 
techniques carry over to the even-parity sector.  To begin, as shown in 
\cite{OsbuETC14} it is straightforward to ensure that the condition number 
of the Wronskian matrix is unity at the start of each numerical integration 
for homogeneous modes by using a QR-preconditioning technique.  The key to 
then limiting the growth of the condition number during numerical 
integration is to begin the integrations as close to the source region as 
possible, while maintaining required accuracy in the initial conditions. 

In the odd-parity sector, for a given multipole and frequency mode, the 
field equations involve two coupled fields.  We can represent the homogeneous 
solutions with a vector $\bscal{\ti{B}}$.  Asymptotically, as 
$r\rightarrow\infty$, the retarded radial fields have a dependence on 
$e^{i\omega r_*}$, where $r_*$ is the usual tortoise coordinate defined by 
$dr_*/dr = f^{-1}$.  In order to place boundary conditions for our numerical 
scheme at a finite radius, $r_\text{out}$, we usually make an asymptotic 
expansion of the fields of the form
\begin{align}
	\label{eqn:oddInfinity}
	\bscal{\ti B}^{\text{Asym}}_j(r_\text{out}) &= e^{i\o r_{*\text{out}}}
	\sum_{s=0}^{s_{\text{max}}} 
	\left[\begin{array}{c}
		\a^{(0)}_{j,s} \\
		\a^{(1)}_{j,s}
	\end{array}\right] 
	\rho_\text{out}^s ,
	\\
	\rho_\text{out} &\equiv \l \o r_\text{out} \r^{-1} , 
\end{align}
where $j=0,1$ and $r_{*\text{out}}\equiv r_*(r_\text{out})$.  This asymptotic 
expansion has limited use unless evaluated in the wave-zone 
$r_\text{out}\gg |\o|^{-1}$.  At low frequencies, a long numerical 
integration is required to reach the source region.  As an alternative to 
this standard approach, we attempted use of an expansion based on the DPA
\begin{align}
	\label{eqn:oddDPA}
	\bscal{\ti B}^{\text{DPA}}_j(r_\text{out})
	&= 
	e^{i\o r_{*\text{out}}} 
	\left[\begin{array}{c} 
		A^{(0)}_{j}(r_\text{out}) \\
		A^{(1)}_{j}(r_\text{out})
	\end{array}\right]
	\\
	A^{(i)}_{j}(r_\text{out}) &\equiv \dfrac{\displaystyle \sum_{s=0}^{s_{\text{max}}/2} 
	\b^{(i)}_{j,s}\,\rho_\text{out}^s}{1+\displaystyle \sum_{s=1}^{s_{\text{max}}/2} \g^{(i)}_{j,s}\,\rho_\text{out}^s}
\end{align}
where $s_{\text{max}}$ is assumed to be even.  It is straightforward to 
compute the DPA coefficients $\b^{(i)}_{j,s}$ and $\g^{(i)}_{j,s}$ from the 
asymptotic expansion coefficients $\a^{(i)}_{j,s}$ (see for example 
Ref.~\cite{PresETC93}).  Quad-precision arithmetic is used to compute the DPA 
coefficients because the linear systems that describe them become increasingly
ill-conditioned when $s_{\text{max}}$ is taken to be large.  This minor use 
of quad-precision in setting boundary conditions is of minimal computational 
cost compared to the previous use of quad-precision in the ODE integrations 
described in Ref.~\cite{OsbuETC14}.  The benefit of the DPA is that the 
boundary conditions can be set at an $r_\text{out}$ that is much closer to 
the source region, thus allowing rapid machine-precision numerical 
integration of the ODEs.

\begin{figure}
\includegraphics[scale=1]{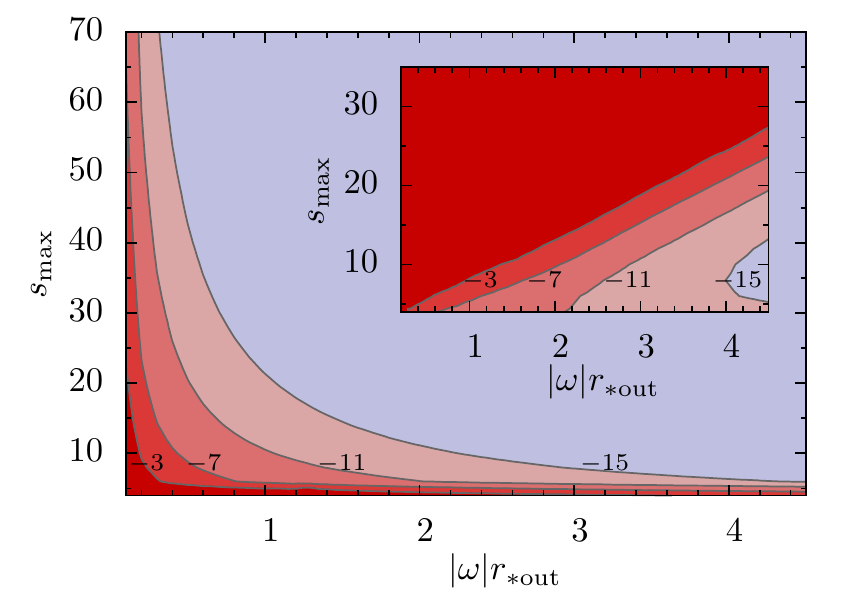}
\caption{
The effectiveness of the diagonal Pade approximant (DPA) method for 
constructing boundary conditions to the homogeneous Lorenz-gauge field 
equations compared to the standard asymptotic expansion.  We calculated the 
relative error for each basis of homogeneous solutions with both methods and 
the worst case is reported.  The contours are of constant relative error 
with $\log_{10}$ scaling, and are given as a function of the number of 
expansion terms $s_\text{max}$ and the location of $r_{*\text{out}}$.  The 
larger plot shows the relative error of the DPA while the inset shows the 
relative error of the asymptotic expansion.  It is apparent that the DPA 
allows initial conditions to be given at approximately a factor of $10$ 
closer to the source region than the asymptotic expansion.  This reduces the 
computation time and improves accuracy by limiting the growth of the 
condition number.  Here we consider the odd-parity case of 
$(l,\o)=(2,10^{-4} M^{-1})$.  Similar results are observed for the 
even-parity sector.
\label{fig:pade}}
\end{figure}

There is no \emph{a priori} guarantee that the DPA will be an improvement over the 
standard asymptotic expansion, but we have tested its validity numerically.  
One appropriate numerical test for how well $\bscal{\ti B}^{\text{DPA}}_j$ 
or $\bscal{\ti B}^{\text{Asym}}_j$ satisfies Eq.~\eqref{eqn:LGeqns} at a 
given $r$ is to use the relevant expansion as initial conditions, perform a 
numerical integration of distance $\D r\sim |\o^{-1}|$, and compare the 
result to the expansion reevaluated at $r+\D r$.  We show in 
Fig.~\ref{fig:pade} that the DPA allows initial conditions to be given at 
approximately a factor of $10$ closer to the source region than the standard 
asymptotic expansion.
The decreased integration distance and reduced rise of condition number allow 
frequencies as small as $|\o M|\ge 10^{-5}$ to be included, which we found 
sufficient for an accurate exploration of the astrophysically relevant 
portion of orbital parameter space.

\subsection{Hybrid self-force}
\label{sec:hyb}

Ideally the Lorenz-gauge self-force code would be used to precompute forces 
that would drive the inspiral via the osculating element method presented in 
Sec.~\ref{sec:osculate}.  Unfortunately, and despite best efforts, at high 
eccentricities the present numerical implementation in the Lorenz-gauge code 
fails to attain the required accuracy in all parts of self-force, as outlined 
in Table \ref{table:required_accuraries}.  The drop in accuracy for orbits with
$e \gtrsim 0.5$ stems from the need to compute and sum over many tens of 
thousands of Fourier-harmonic modes.  

Fortunately, it is not necessary to know all parts of the self-force with 
equal accuracy (again see Table \ref{table:required_accuraries}).  The most 
critical accuracy requirement is on the adiabatic, orbit-averaged part of the 
dissipative self-force.  This part of the self-force can be determined from 
energy and angular momentum fluxes at infinity and the horizon, and does not 
require a local calculation.  We can obtain the fluxes from the Lorenz-gauge 
code or from a separate RWZ code.  This is the basis of the hybrid scheme 
outlined previously in \cite{OsbuETC14}, which augments the Lorenz-gauge 
results with highly accurate flux data from a RWZ code.  In this section we 
review how to construct such a ``hybrid self-force,'' which is sufficiently 
accurate to compute inspirals with phase error less than $0.1$ radians.

It begins by noting that for a background geodesic $u_t=-\mathcal{E}$ and 
$u_\varphi=\mathcal{L}$ are constants of the motion, and thus the covariant 
form of Eq.~\eqref{eqn:forced} for the $t$ and $\varphi$ components will 
determine gradual changes in the particle's specific energy and angular 
momentum.  Multiplying by $\mu$ we get rates of change with respect to proper 
time of the particle's energy and angular momentum.  Integrating these over 
proper time to find averages, the orbit-averaged rate of gain (or loss) of 
energy and angular momentum with respect to coordinate time due to the 
self-force is 
\begin{align}
	&\mu \langle\dot{\mathcal{E}}\rangle = -\frac{1}{T_r}\int_0^{\mathcal{T}_r} F_t \, d\tau 
	= -\frac{\mathcal{T}_r}{T_r} \langle F_t \rangle_\tau ,
	\\
	&\mu \langle\dot{\mathcal{L}}\rangle = \frac{1}{T_r}\int_0^{\mathcal{T}_r} F_\varphi \, d\tau 
	= \frac{\mathcal{T}_r}{T_r} \langle F_\varphi \rangle_\tau .
\end{align}
In these expressions the overdot indicates differentiation with respect to 
coordinate time, $t$, angle brackets with a $\tau$ subscript indicate a 
proper-time average, and angle brackets with no subscript indicate a 
coordinate time average.  The rate at which the particle loses energy and 
angular momentum must be balanced by the averaged asymptotic fluxes.  This 
balance gives
\begin{align}
\label{eqn:ELdot}
\mu \langle\dot{\mathcal{E}}\rangle = - \langle\dot{E}\rangle, \q\q\q\q 
\mu \langle\dot{\mathcal{L}}\rangle = - \langle\dot{L}\rangle,
\end{align}
where $\langle\dot{E}\rangle$ and $\langle\dot{L}\rangle$ are the average 
rates at which energy and angular momentum are radiated, respectively.  These
balance formulas can then be related to the adiabatic self-force components via
\begin{align}
\label{eqn:adF}
F^\text{ad}_t 
= \langle F_t \rangle_\tau 
= \frac{T_r}{\mathcal{T}_r}\langle\dot{E}\rangle ,
\q\q F^\text{ad}_\varphi 
= \langle F_\varphi \rangle_\tau 
= -\frac{T_r}{\mathcal{T}_r}\langle\dot{L}\rangle .
\end{align}
Harking back to our discussion in Sec.~\ref{sec:effects_of_SF}, the hybrid 
method settles on adopting the average over proper time to define the 
adiabatic part of the self-force.

The object then is to remove $F_{t/\varphi}^\text{ad}$ from our Lorenz-gauge self-force 
and replace it with the values computed at much higher accuracy with our 
RWZ code.  We can separate out the adiabatic component of the self-force from 
our numerical Lorenz-gauge results by noting that the oscillatory part of the 
self-force averages to zero over an orbital period.  This motives a Fourier 
decomposition of the self-force in proper time,
\begin{align}
\label{eq:FalphaFourier}
&F_\alpha = \ti{a}^{(\alpha)}_0 
+ \sum_{n=1}^\infty\big[ \ti{a}^{(\alpha)}_n \cos(2\pi n\tau/\mathcal{T}_r) 
+ \ti{b}^{(\alpha)}_n \sin(2\pi n\tau/\mathcal{T}_r) \big] ,
\end{align}
where $\alpha = \{t,\varphi\}$ (we address the radial component of the 
self-force momentarily).  Comparing to Eq.~\eqref{eq:F_ad_and_osc} we see that
\begin{align}
\label{eqn:oscFalpha}
F^\text{ad}_\alpha &= \ti{a}^{(\alpha)}_0	\\
F^\text{osc}_\alpha 
&= \sum_{n=1}^\infty\big[ \ti{a}^{(\alpha)}_n \cos(2\pi n\tau/\mathcal{T}_r) 
+ \ti{b}^{(\alpha)}_n \sin(2\pi n\tau/\mathcal{T}_r) \big] .
\end{align}
The ingredients are now at hand and we construct the hybrid self-force via
\begin{align}
\label{eq:F_hyb}
F^\text{hyb}_\alpha(p,e,v) 
= F^\text{ad(RWZ)}_\alpha(p,e) + F_\alpha^\text{osc(Lor)}(p,e,v) ,
\end{align}
with explicit dependence on orbital parameters indicated.

In computing $F^\text{ad(RWZ)}_\alpha$ we use a RWZ code based off of 
Refs.~\cite{HoppEvan10,HoppETC15}.  In constructing $F_\alpha^\text{osc(Lor)}$ 
we make use of the discrete Fourier transform (DFT) to compute the amplitudes 
in Eq.~\eqref{eq:FalphaFourier} (see Ref.~\cite{HoppETC15} where these 
techniques are used in a similar application).  The algorithmic roadmap for 
constructing $F^\text{hyb}_\alpha$ for a given $(p,e)$ is then the following:
\begin{enumerate}

\item
\emph{Compute Lorenz gauge self-force.--}
See subsection \ref{sec:mp}.  Our code is configured to return the 
contravariant components of the self-force $F^\a$ at a large number of time 
samples equally spaced in $v$.  We construct the covariant self-force at the 
same $v$ samples by lowering the index using the background metric.

\item
\emph{Interpolate $F_\alpha(v)$ using DFT.--}
Compute the coefficients $\ti{g}^{(\a)}_n$ and $\ti{h}^{(\a)}_n$ of the 
Fourier series expansion 
\begin{align}
	\label{eqn:vFourier}
	F_\a = \sum_{n=0}^N \big[ \ti{g}^{(\a)}_n \cos(nv) + \ti{h}^{(\a)}_n \sin(nv) \big]
\end{align}
using a DFT applied to the equally spaced-in-$v$ numerical data.  
Equation~\eqref{eqn:vFourier} can then be used to construct $F_\alpha(v)$ at 
arbitrary values of $v$.

\item
\emph{Compute list of $v$ values consistent with equal $\tau$ spacing.--}
Special functions \cite{FujiHiki09} or root finding of the Fourier 
representation of $\tau(v)$ \cite{HoppETC15} can be used to choose a list of 
equally spaced $\tau$ values and find the corresponding list of $v$ values.  
We use the root finding method.

\item
\emph{Compute $\tau$-Fourier series of $F_\alpha$.--}
Construct the equally spaced-in-$\tau$ values of $F_\alpha$ using the $v$ 
values from the previous step and interpolating using Eq.~\eqref{eqn:vFourier}.
The DFT of this data gives the desired Fourier amplitudes 
$\ti{a}^{(t/\varphi)}_n$ and $\ti{b}^{(t/\varphi)}_n$ in 
Eq.~\eqref{eq:FalphaFourier}.  A strong check is to compare $\ti{a}^{(t)}_0$ 
with $\langle\dot{E}\rangle^\text{RWZ}T_r/\mathcal{T}_r$ and 
$\ti{a}^{(\varphi)}_0$ with 
$-\langle\dot{L}\rangle^\text{RWZ}T_r/\mathcal{T}_r$.  These should agree to 
as many digits as are attainable from the Lorenz gauge results (see, for 
example, Table V of Ref.~\cite{OsbuETC14}).

\item
\emph{Construct hybrid force.--}
The hybrid force is constructed using Eq.~\eqref{eq:F_hyb}.  The adiabatic 
piece is computed with the RWZ fluxes using Eq.~\eqref{eqn:adF}.  The 
oscillatory part is computed using the Fourier coefficients from the previous 
step with Eq.~\eqref{eqn:oscFalpha}.

\item
\emph{Construct contravariant hybrid force with equal $v$ spacing.--}
Our osculating elements scheme is formulated with the contravariant 
components; therefore, we raise the index with the background metric.  Note 
that this causes $F_\text{ad}^\a$ to vary over an orbit.  In the section 
that follows, we interpolate over the $(p,e)$ parameter space and find it 
convenient to resample with equal $v$ spacing.

\end{enumerate}

So far we have ignored hybridization of the $r$ component of the self-force.  
In principle $F^r_\text{hyb}$ could be constructed from the orthogonality 
condition $F^\a_\text{hyb} u_\a = 0$.  Instead of doing so, we express the 
$e$ and $p$ evolution equations in terms of only $F^t_\text{hyb}$ and 
$F^\varphi_\text{hyb}$, eliminating the need of the $r$ component of the 
self-force in those two equations.  There remains the equation for $\chi_0$ 
evolution.  Rewriting that equation in terms of the $t$ and $\varphi$ 
components of the self-force is not numerically practical as it introduces a division by $u^r$ which is zero at the orbital turning points.  
Fortunately, in the $\chi_0$ evolution the conservative part of the 
self-force dominates over the dissipative part by a factor of $\epsilon$ 
(see Sec.~\ref{sec:hyb_performance} and Fig.~\ref{fig:chi0err}). 
Since hybridization only (subtly) alters the dissipative part, hybridization 
would affect the evolution of $\chi_0$ at a level many orders of magnitude 
below the dominant behavior.

\section{Interpolation of the hybrid self-force across the $(p,e)$ parameter 
space}
\label{sec:interp}

In order to numerically integrate the osculating element equations 
\eqref{eqn:dedchi}-\eqref{eqn:dchi0dchi} we need to supply the self-force at 
arbitrary values of $(p,e,v)$.  Whilst our Lorenz-gauge code is capable of 
rapidly computing the self-force, it is not sufficiently quick to allow it 
to be directly coupled to the integration of the osculating elements.  
Instead we populate the relevant portion of the $(p,e)$ parameter space with 
a few thousand data points and interpolate to the intervening values.  This 
section describes our interpolation procedure.

\subsection{Sampling the hybrid self-force}

Equally sampling the hybrid self-force in $(p,e)$ space is not optimal, 
especially near the separatrix where small changes in $p$ can lead to large 
changes in the value of the self-force.  The behavior of the radiated fluxes 
near the separatrix \cite{CutlKennPois94} suggests that a good 
parametrization in this region is $y(x) \sim -1/\ln x$, where 
$x\equiv p-2e-6$.  However, this choice is not well suited to points away 
from the separatrix so we construct a function that smoothly transitions 
$y(x)$ to be proportional to $x$ away from the separatrix
\begin{align}
\label{eqn:y}
	y(x) &\equiv \Bigg\{ \begin{array}{c} (x+8)w(x,6)-\dfrac{35[1-w(x,6)]}{\ln\left(x/80\right)} 
	, \q x<6 \\ x+8, \q x \ge 6 \q\q\q\q\q\q\q \end{array} ,
\end{align}
where $w(x,d)$ is a smooth transition function of width $d$ given by
\begin{align}
w(x,d) &\equiv \frac{1}{2}+\frac{1}{2}\tanh\left[\tan\left( \frac{\pi x}{2d} \right)
 - \cot\left( \frac{\pi x}{2d} \right) \right].
\end{align}

\begin{figure*}
\center
\includegraphics[width=\textwidth]{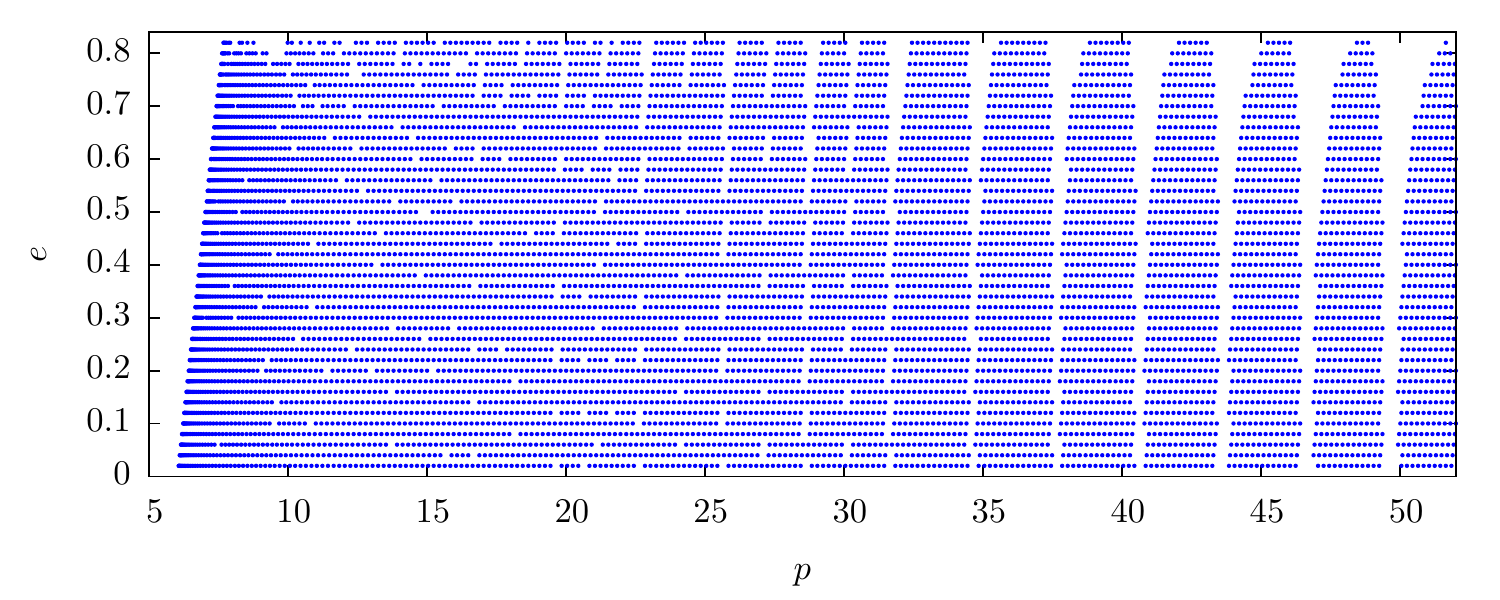}
\caption{
Data used for interpolation of the oscillatory self-force.  We computed data 
for 9602 unique orbits at a cost of 2308 CPU hours.  The adiabatic data are  
computed over approximately the same domain, but with four times the density 
and no gaps due to orbital resonances.  Explicitly, we computed adiabatic 
data for 43875 unique orbits at a cost of 2054 CPU hours.  Most of the gaps 
in the data set correspond to orbital resonances where small (nonzero) 
Fourier-mode frequencies are encountered (these modes are difficult for our 
Lorenz-gauge code to compute \cite{OsbuETC14}). 
\label{fig:oscgrid}
}
\end{figure*}

We computed the adiabatic part of the self-force using the RWZ gauge code 
on a grid with $\D y=0.1$, $y_\text{min}=4$, $y_\text{max}=59$, $\D e=0.01$, 
$e_\text{min}=0.01$, and $e_\text{max}=0.83$.  We computed the oscillatory 
part of the self-force (and the full self-force $F^\a$) using the Lorenz 
gauge code on a grid with $\D y=0.2$, $y_\text{min}=4.4$, $y_\text{max}=59$, 
$\D e=0.02$, $e_\text{min}=0.02$, and $e_\text{max}=0.82$ (see 
Fig.~\ref{fig:oscgrid}).  There are some gaps in the data, especially in the 
oscillatory part where we avoid orbits with nonzero frequencies $\o_{mn}$ 
smaller than $10^{-5}M^{-1}$.

For the adiabatic self-force we computed data for 43875 unique orbits at a 
cost of 2054 CPU hours.  For the oscillatory, Lorenz-gauge self-force we 
computed data for 9602 unique orbits at a cost of 2308 CPU hours.  We also 
explored spacing the data using a reduced order model \cite{Field:2013cfa}.  
Our initial tests suggested this would be a promising method to reduce the 
computational burden but we did not pursue if further.  Such methods might 
be important though when interpolating the self-force over the larger 
parameter space of geodesics in Kerr spacetime.

\subsection{Interpolation of the self-force}

The periodicity of the geodesic self-force suggests using a Fourier series for 
interpolation in time \cite{WarbETC12}
\begin{align}
	\label{eqn:vinterp}
	F^\a &= \mu^2 \sum_{n=0}^{n_\text{max}} \left[ a^\a_n(e,y) \cos (nv)+b^\a_n(e,y) \sin (nv) \right] .
\end{align}
The Fourier coefficients $a^\a_n(e,y)$ and $b^\a_n(e,y)$ can then be 
interpolated across orbital parameter space ($e$ and $y$).  We truncate the 
Fourier series at $n_\text{max}=13$ because we have found that to be a 
sufficient number of harmonics to represent the force at our accuracy goals.
Our self-force codes output the Fourier amplitudes $a^\a_n$ and $b^\a_n$ 
directly by computing the DFT of data with a large number of equally spaced 
$v$ samples.  Note that for the adiabatic part $b^\a_n=0$.  As an example we 
will consider the interpolation of $a^\a_n$, but the same techniques apply 
to $b^\a_n$.  We separately interpolate the Fourier amplitudes of the 
adiabatic, oscillatory, and nonhybrid parts of the self-force.

A similar method was used by Ref.~\cite{WarbETC12} to interpolate the 
(nonhybrid) self-force.  In that work they interpolated over a parameter 
space spanning $6+2e<p<12$ and $0\le e \le0.2$ by performing global fits to 
power series in $p$ and $e$.  Global fits are challenging to work as the fit 
has to incorporate the post-Newtonian-like behavior of the self-force in the 
weak field as well as the behavior in the strong-field using a small set of 
parameters.  As such the fidelity of the model is reduced.  In this work we 
use a local fitting procedure.  This results in a great deal more parameters 
that describe how the self-force varies over the parameter space, but in 
exchange the fidelity of our interpolation model is greatly improved.  In 
fact, the accuracy of our model is within an order of magnitude of the 
underlying data.

\begin{figure}
\includegraphics[scale=.67]{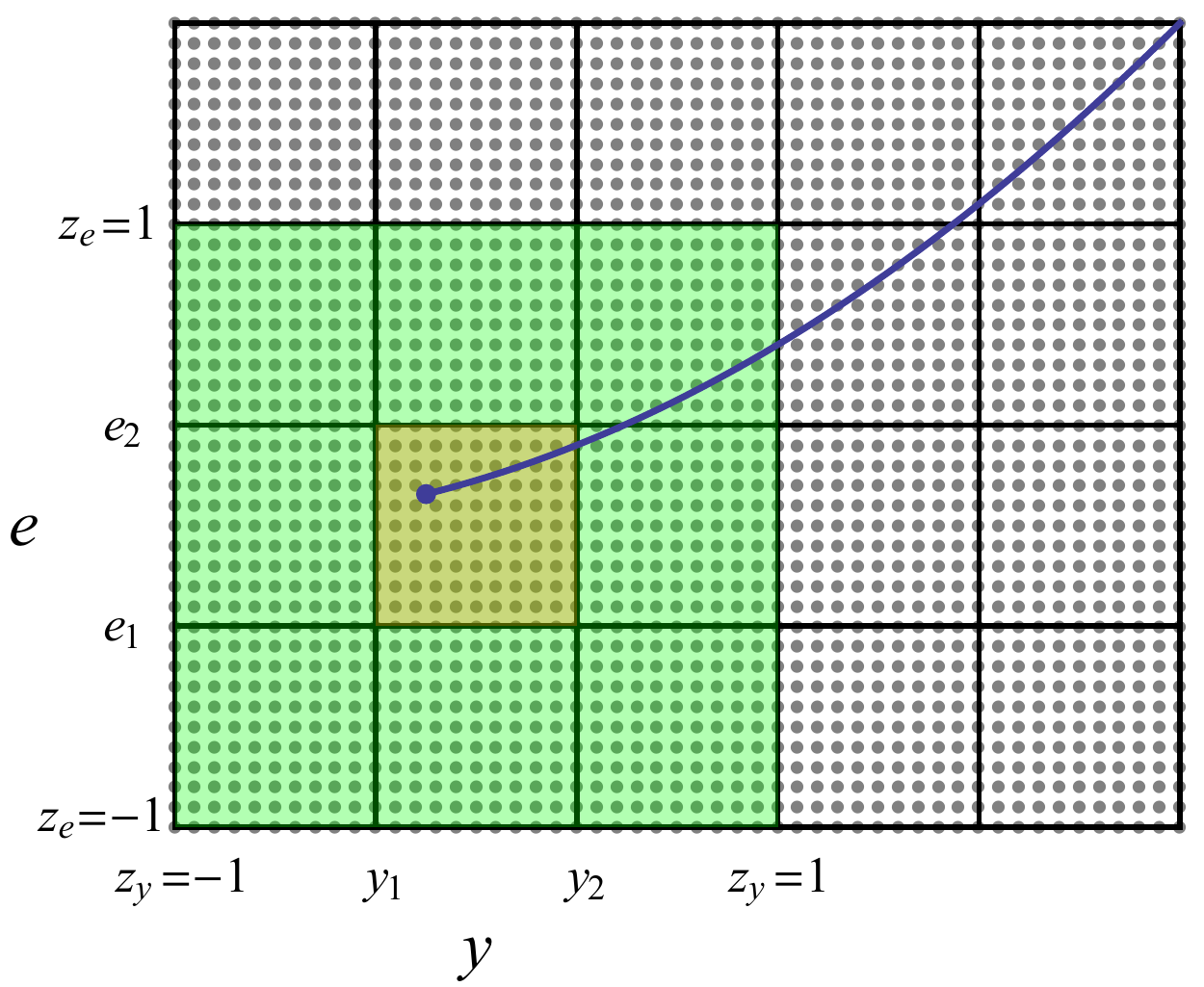}
\caption{The local discretization used for interpolation over the $(e,y)$ 
parameter space (see Eq.~\eqref{eqn:y} for the defintion of $y$).  The blue 
line represents the inspiral trajectory with a point at the current position.  
The yellow zone is the inspiral's current subdomain.  The interpolation is 
performed with data (gray dots) from the yellow and green zones.
\label{fig:discretize}}
\end{figure}

Our local interpolation scheme begins by subdividing the domain into a grid 
of smaller rectangular zones.  To obtain the self-force in a particular zone 
(with domain $e_1\le e<e_2$, $y_1\le y<y_2$) we interpolate using data from 
the nearest 9 zones (all the surrounding rectangles including the current 
one; see Fig.~\ref{fig:discretize}).  We rescale $e$ and $y$ into new 
variables $z_e$ and $z_y$ that equal $-1$ at the leading edge of the 
interpolation region and $+1$ at the trailing edge of the interpolation 
region.  We then make a Chebyshev interpolation
\begin{align}
	&z_e \equiv \frac{2e-e_2-e_1}{3(e_2-e_1)} , \q\q z_y \equiv \frac{2y-y_2-y_1}{3(y_2-y_1)} ,
	\\
	&\q\q a^\a_n = \sum_{i=0}^{i_\text{max}}\sum_{j=0}^{j_\text{max}} 
	\s^\a_{nij} T_i(z_e)T_j(z_y) , \label{eqn:cheby}
\end{align}
where $T_i(z)$ is the Chebyshev polynomial of the first kind.  To ensure the 
correct units, $\s^t_{nij}$ and $\s^r_{nij}$ are implied to have overall 
factors of $M^{-2}$ while $\s^\varphi_{nij}$ is implied to have an overall 
factor of $M^{-3}$.  We evaluate Eq.~\eqref{eqn:cheby} for every data point 
in the interpolation region, which is a linear system for the unknown 
coefficients $\s^\a_{nij}$.  We require that the number of equations be 
greater than the number of unknowns, or equivalently that the number of data 
points is greater than $(i_\text{max}+1) \times (j_\text{max}+1)$.  We use 
least-squares fitting to compute $\s^\a_{nij}$.  This fit is precomputed for 
every subdomain to facilitate rapid numerical evaluation.  Once the 
interpolation coefficients $\s^\a_{nij}$ are known for each subdomain 
Eqs.~\eqref{eqn:vinterp} and \eqref{eqn:cheby} give the interpolated 
self-force. 

\begin{figure}
\includegraphics[scale=1]{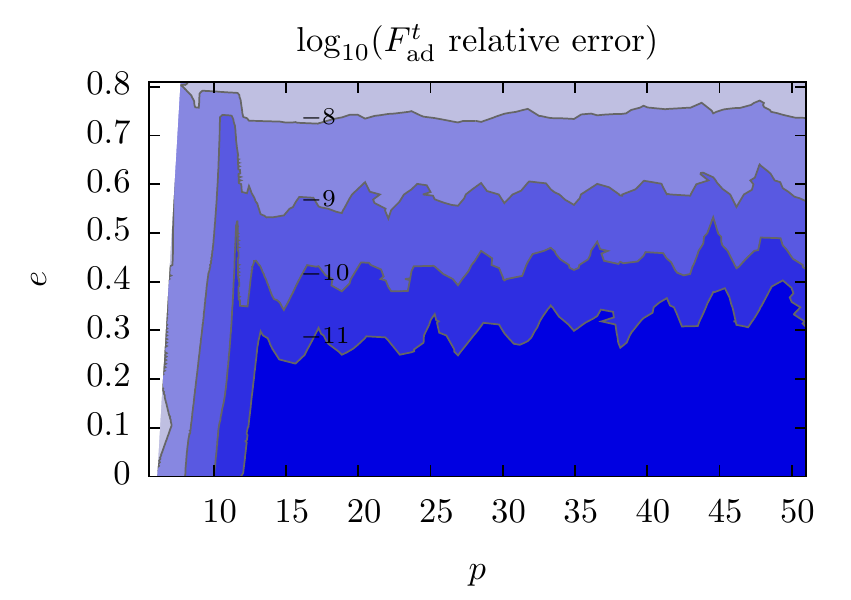}
\caption{Estimates of interpolation error in adiabatic part of self-force.  
We estimate the interpolation error of $F^t_\text{ad}$ by computing orbits 
independent of those used for fitting interpolation coefficients and 
comparing with interpolated self-force values.  The interpolation model 
recovers $F^t_\text{ad}$ across parameter space with an error no worse 
than $\sim 10^{-8}$ (better for lower eccentricities and away from the 
separatrix). Similar results are observed for the other components of the self-force. 
\label{fig:adiabaterr}}
\end{figure}

\begin{figure}
\includegraphics[scale=1]{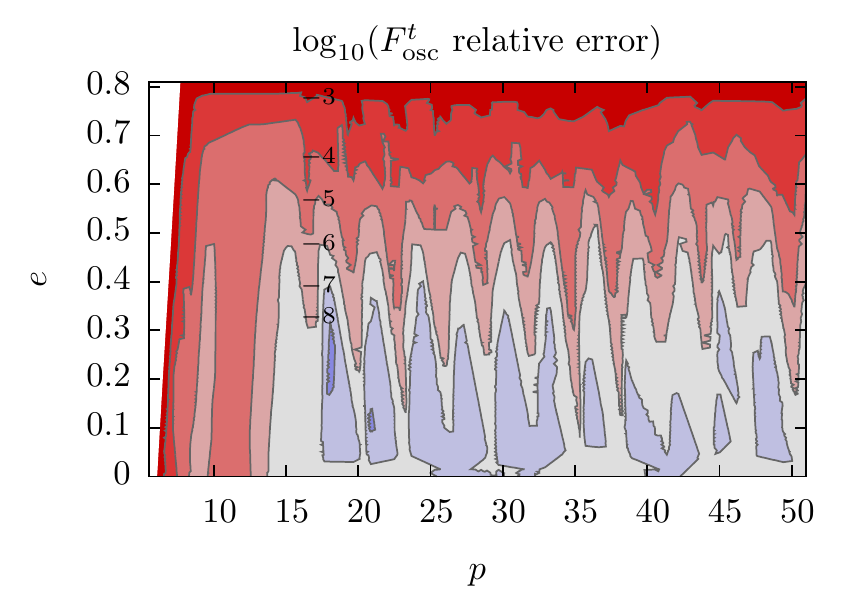}
\caption{Estimates of interpolation error in oscillatory part of self-force.  
We estimate the interpolation error of $F^t_\text{osc}$ by computing orbits 
independent of those used for fitting interpolation coefficients and 
comparing with interpolated self-force values.  The interpolation model 
recovers $F^t_\text{osc}$ across parameter space with an error no worse than 
$\sim10^{-3}$ (better for lower eccentricities and away from the separatrix).
The larger error at high eccentricity is a limitation of the underlying data 
from the Lorenz-gauge code and motivates the hybrid scheme. Similar results are observed for the other components of the self-force.
\label{fig:oscerr}}
\end{figure}

For the adiabatic self-force interpolation we use 145 $y$-zones, 20 $e$-zones, 
and take $i_\text{max}=j_\text{max}=12$.  For the oscillatory (and nonhybrid) 
self-force interpolation we use 72 $y$-zones, 10 $e$-zones, and we take 
$i_\text{max}=j_\text{max}=10$.  To check the accuracy of the interpolation 
we compute the self-force for orbits not used in the fit for interpolation 
coefficients and compare with the interpolated result (see 
Figs.~\ref{fig:adiabaterr} and \ref{fig:oscerr}). 

\section{Initial conditions with matched frequencies}
\label{sec:freq_match}

A number of works have argued that the gravitational waveforms from inspirals 
computed using only the adiabatic self-force will be sufficient for detection 
with space-based detectors \cite{Gair:2004iv,HughETC05,Huerta:2008gb}.  The 
best way to assess this claim is by comparing inspirals computed with and 
without the oscillatory and conservative self-force corrections.  The question 
then arises of how should one compare two such inspirals?  In answering this 
question it is important to remember that an adiabatic inspiral and an 
inspiral computed using the full self-force live in two different 
spacetimes\footnote{for instance, by artificially turning off the conservative 
self-force, one is excluding the conservative part of the metric perturbation} 
so that a direct coordinate comparison (say by setting the initial 
$p,e,\chi_0,\varphi_p,t_p$ the same for both inspirals) is not ideal.  A more 
appropriate comparison can be made by choosing the initial (gauge invariant) 
frequencies of the orbit to be the same.  The utility of adiabatic self-force 
inspirals can then be assessed by comparing the accumulated azimuthal phase 
with that from an inspiral computed using the full self-force.

Before we consider how to match the initial inspiral frequencies in practice 
let us briefly discuss why simply matching the values of $p$ and $e$ for each 
inspiral is not optimal.  The goal is to provide initial conditions for the 
approximate inspiral that minimises the phase difference with an inspiral 
computed using the full self-force.  For clarity we'll consider the case for 
quasicircular orbits where there is only one orbital frequency.  For each 
inspiral we can expand the phase evolution in a Taylor series about $t_p=0$ 
and write the difference between the two inspirals as:
\begin{align}
\label{eq:phase_diff}
\Delta\varphi_p(t_p) 
= (\varphi_0^\text{apx} - \varphi_0^\text{full}) 
+ \left(\Omega_\varphi^\text{apx} 
- \Omega_\varphi^\text{full} \right)t_p + \mathcal{O}(t_p^2) ,
\end{align}
where an "apx" superscript denotes a quantity associated with the inspiral 
that is computed using an approximation to the full self-force.  Examples of 
such approximations are the adiabatic approximation, calculated by flux 
balance arguments, or the dissipative approximation, which excludes the 
conservative effects but retains the oscillatory dissipative self-force.  
The "full" superscript denotes a quantity associated with an inspiral 
computed using the full self-force.  Without loss of generality we can set 
$\varphi_0^\text{apx} = \varphi_0^\text{full}$ and then from 
Eq.~\eqref{eq:phase_diff} we see that equating the two initial orbital 
frequencies will remove the initial linear growth in the phase difference. With the frequencies matched the phase difference will still grow in time, but at the slower quadratic rate.

In order to match the initial frequencies for an eccentric inspiral we must 
find values of $p_0$ and $e_0$ for each inspiral such that
\begin{align}
\Omega_\varphi^\text{apx}(p_0^\text{apx},e_0^\text{apx}) 
- \Omega_\varphi^\text{full}(p_0^\text{full},e_0^\text{full}) = 0 \\
\Omega_r^\text{apx}(p_0^\text{apx},e_0^\text{apx}) 
- \Omega_r^\text{full}(p_0^\text{full},e_0^\text{full}) = 0 .
\end{align}
In general, setting $p_0^\text{apx} = p_0^\text{full}$ and 
$e_0^\text{apx} = e_0^\text{full}$ will not match the frequencies.  Instead, 
we match the frequencies using the following procedure.  We choose values for 
$p_0^\text{full}$ and $e_0^\text{full}$, calculate 
$\Omega_\varphi^\text{full}$ and $\Omega_r^\text{full}$, and then use a root 
finding algorithm to find the values of $p_0^\text{apx}$ and $e_0^\text{apx}$ 
that gives the same value for the initial frequencies for the approximate 
inspiral.  It is interesting to note that the relation between the orbital 
frequencies and ($p,e$) is not one-to-one for orbits near the separatrix in 
the $(p,e)$ parameter space \cite{Warburton:2013yj}.  Nonetheless, so long as 
the frequency matching is performed far from the separatrix, as is always the case 
in this work, there is no ambiguity in matching the frequencies.

Calculating the orbital frequencies including the self-force corrections is 
achieved by integrating the osculating orbit equations over one orbital 
period.  Explicitly, we change the integration variable from $\chi$ to 
$v = \chi - \chi_0$ (using $dv/d\chi = 1-d\chi_0/d\chi$) in 
Eqs.~\eqref{eqn:t},\eqref{eqn:phi},\eqref{eqn:dedchi}-\eqref{eqn:dchi0dchi} 
and integrate the equations from $v=0$ to $v=2\pi$, using the relevant 
approximation to the self-force in 
Eqs.~\eqref{eqn:dedchi}-\eqref{eqn:dchi0dchi}.  The time elapsed and 
azimuthal phase accumulated between periastron passage we denote by $T_r$ 
and $\Delta\varphi$.  Equations \eqref{eq:freqs} can then be used to compute 
the associated frequencies.

\section{Highly-eccentric inspiral results} 
\label{sec:res}

In this section we present our main results--a sample of inspirals computed 
using our hybrid geodesic self-force inspiral model.  The physical results 
for extreme- and intermediate-mass-ratio inspirals are presented in 
Sec.~\ref{sec:EMRI}.  First, though, we quantify the performance of our 
hybrid self-force method.

\subsection{Performance of the hybrid self-force method}
\label{sec:hyb_performance}

Our hybrid scheme aims to produce a self-force that is sufficiently accurate 
to capture the leading and subleading contributions to the inspiral phase 
from the first-order-in-the-mass-ratio self-force.  As discussed earlier, the 
raw self-force output from the Lorenz-gauge code does not meet this 
requirement for all eccentricities, and so we supplement those results with 
high-accuracy flux data from a RWZ code (see Sec.~\ref{sec:hyb}).

To test whether our hybrid method allows the accumulated phase of an inspiral 
to be tracked to within $\sim 0.1$ radians, we perform several sensitivity 
tests.  The sensitivity of the inspiral phase to a relative error $\d$ in the 
oscillatory part of the self-force is tested by computing two inspirals, one 
where we introduce a uniform positive perturbation (trial error) 
$F^\a_\text{osc}\rightarrow (1+\d)F^\a_\text{osc}$ and another where we 
introduce a uniform negative perturbation 
$F^\a_\text{osc}\rightarrow (1-\d)F^\a_\text{osc}$.  The absolute response 
in the orbital elements to these introduced errors is estimated by calculating 
the half-difference between the two perturbed inspirals.  With the sensitivity 
to errors in the oscillatory part of the self-force tested, we then make an 
equivalent test on the adiabatic part, $F^\a_\text{ad}$. 

\begin{figure}
\includegraphics[scale=1]{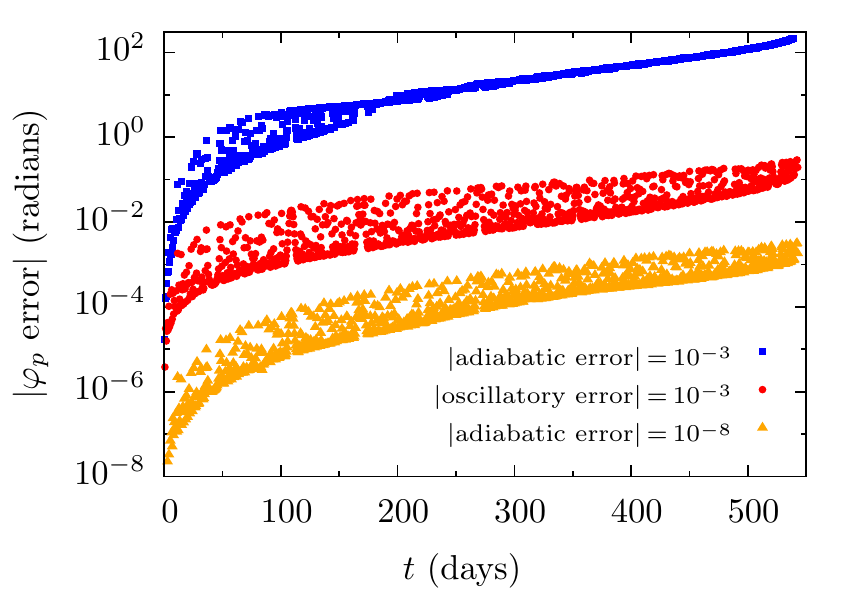}
\caption{Sensitivity of inspiral phase to error in the self-force.  We test the 
sensitivity of the inspiral phase, $\varphi_p$, to errors in $F^\a_\text{ad}$ 
and $F^\a_\text{osc}$ by independently perturbing each part of the self-force 
with uniform errors of the indicated relative size, $\d$.  At a relative size 
of $\d$, the expectation is that trial errors introduced in the adiabatic part 
of the self-force should have an effect that is a factor $\e^{-1}$ larger than 
the effect of comparable errors injected in the oscillatory part of the 
self-force.  The observed ratio is less dramatic but nevertheless indicates 
that computing the adiabatic part more accurately by orders of magnitude is 
crucial.  The inspiral parameters were set to be $e_0=0.7$, $p_0=10$, 
$\chi_{00}=0$, and $\e=10^{-5}$.  The time scale is set by assuming 
$M=10^6 M_\odot$. 
\label{fig:phiSensitivity}}
\end{figure}

Figures~\ref{fig:adiabaterr} and \ref{fig:oscerr} showed previously that the 
adiabatic and oscillatory parts of the self-force in the hybrid scheme are 
accurate to at least $10^{-8}$ and $10^{-3}$, respectively, and to much higher 
accuracy over most of orbital parameter space.  The issue then is whether 
these error levels translate into requisite bounds on phase error.  To 
determine this we ran the error sensitivity tests with error injections at 
these levels.  In 
Fig.~\ref{fig:phiSensitivity}, we perturbed the adiabatic and oscillatory 
components of the self-force with relative errors of $\pm 10^{-8}$ (yellow) 
and $\pm 10^{-3}$ (red), respectively.  We then tracked the relative drift 
in the cumulative azimuthal phase during the inspiral.  For 
$\epsilon=10^{-5}$ we find that a $\d=10^{-8}$ perturbation in 
$F^\a_\text{ad}$ induces a $\sim 10^{-3}$ radian error in $\varphi_p$.  For 
$F^\a_\text{osc}$, a perturbation of $\d=10^{-3}$ causes an absolute error 
of $\sim 0.1$ radians in $\varphi_p$.  We conclude that the numerical accuracy 
of the hybrid self-force model is sufficient to hold phase errors to less 
than 0.1 radians at the highest eccentricities $e \sim 0.7$.  At lower 
eccentricities the inspiral phase error is smaller by orders of magnitude.  
Also indicated in the plot (blue) is the phase drift that would result for a 
$e = 0.7$ inspiral if only the Lorenz-gauge self-force had been used, 
demonstrating clearly the need to isolate the adiabatic part and compute it 
to higher accuracy (i.e., use the hybrid model). 

\begin{figure}
\includegraphics[scale=1]{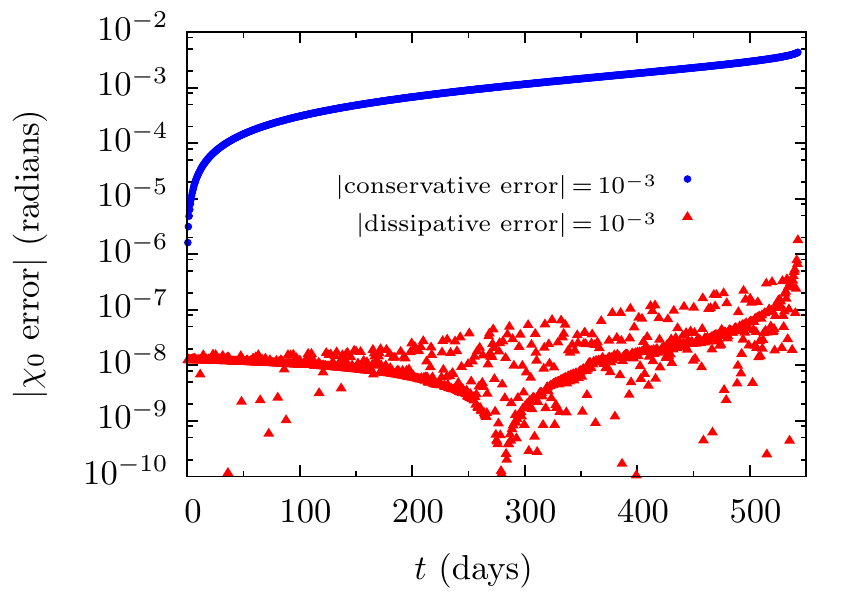}
\caption{Sensitivity in the evolution of $\chi_0$ to error in the radial self-force.  
To test the propagation of errors into $\chi_0$, we perturbed the force 
components in the $\chi_0$ evolution equation while leaving the $e$ and $p$ 
evolution equations unaffected.  Furthermore, we independently introduced 
errors into $F^r_\text{cons}$ and $F^r_\text{diss}$ in the $\chi_0$ equation.  
At the worst case error level, the $F^r_\text{diss}$ clearly has little 
influence on the evolution of $\chi_0$.  Since only the dissipative self-force is 
affected by hybridization, we see that the hybrid force is not essential in the evolution 
of $\chi_0$.  We also see that, at this same error level, the conservative self-force is accurate 
enough to hold errors in $\chi_0$ to 0.01 radians or less.  The inspiral 
parameters were $e_0=0.7$, $p_0=10$, $\chi_{00}=0$, and $\epsilon=10^{-5}$.  
The time scale is set by assuming $M=10^6 M_\odot$. 
\label{fig:chi0err}}
\end{figure}

As discussed in Sec.~\ref{sec:hyb}, within our scheme we create hybrid 
self-force values for $F^t$ and $F^\varphi$ but do not create a hybridized 
$F^r$.  In principle $F^r_\text{hyb}$ could be computed using 
$F^\alpha_\text{hyb} u_\alpha = 0$ but such a construction involves dividing 
by $u^r$, which vanishes at the radial turning points.  Instead we use the 
Lorenz-gauge (nonhybrid) $F^r$ when computing the evolution of $\chi_0$ 
($F^r$ is not directly required to evolve $p$ and $e$ as we write their 
evolution equations in terms of $F^t$ and $F^\varphi$ only).  To ensure that 
using the nonhybrid result for $F^r$ does not adversely affect our results, 
we made a sensitivity test in the evolution of $\chi_0$.  As 
Fig.~\ref{fig:chi0err} indicates, the dissipative part of $F^r$, which is 
the element that would be affected by hybridization, has little influence 
on the evolution of $\chi_0$.  Instead, as the figure shows, it is the 
conservative part of $F^r$ that dominates the evolution of $\chi_0$, and 
our scheme is accurate enough to hold errors in $\chi_0$ to 0.01 radians.  

\subsection{EMRI and IMRI results}\label{sec:EMRI}

\begin{figure}
\includegraphics[width=8.5cm]{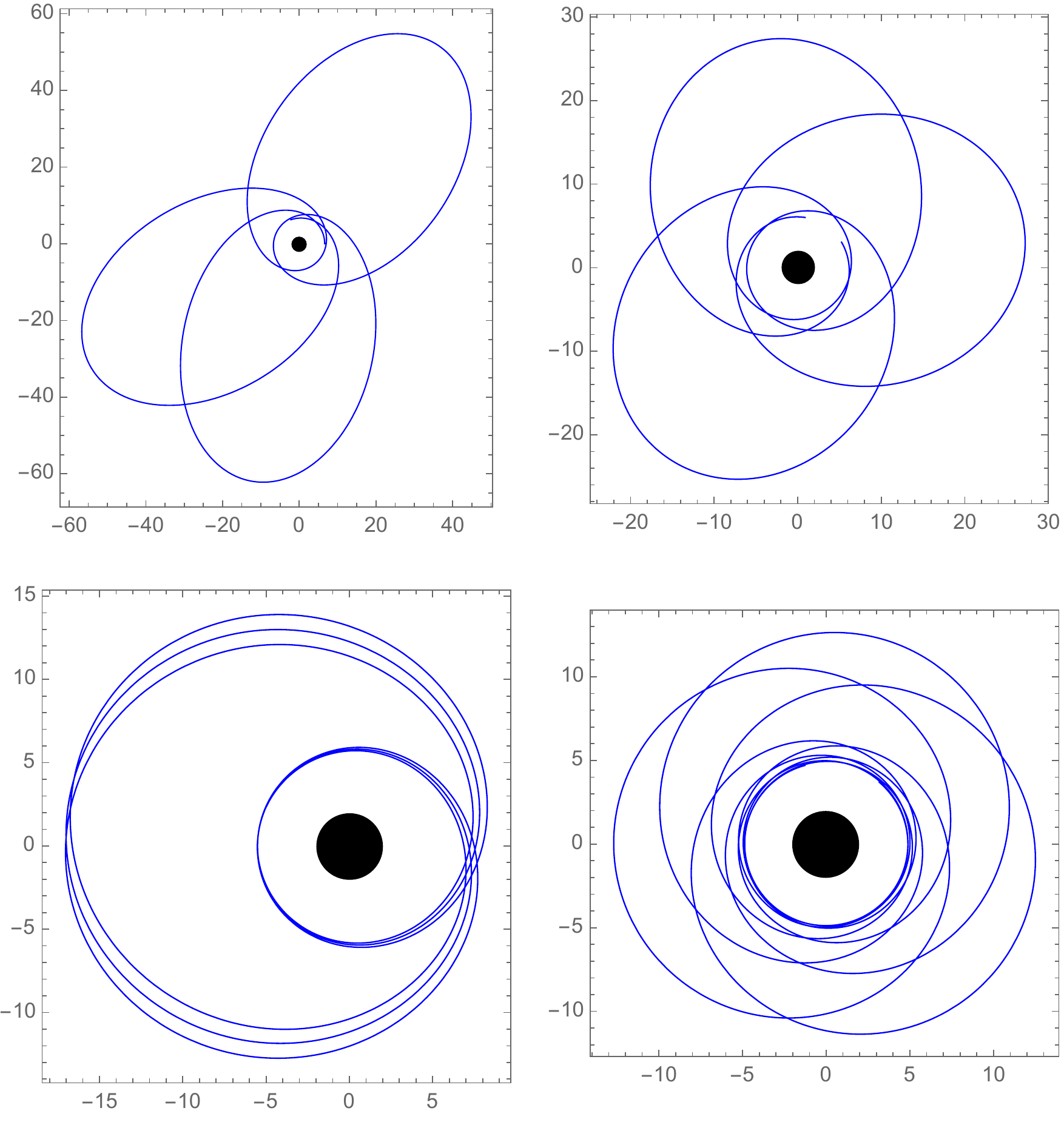}
\caption{Sample snapshots of an inspiral with $M=10^6 M_\odot$ and 
$\epsilon = 10^{-5}$.  The inspiral is plotted in Boyer-Lindquist coordinates 
with $x=(r_p/M) \cos(\varphi_p), y=(r_p/M) \sin(\varphi_p)$.  Each snapshot 
shows three periastron passages of the (counterclockwise moving) inspiral 
and the central black hole is drawn to scale.  The initial configuration is 
$\sim2115.5$ days from plunge and is shown in the top left panel.  The initial 
parameters are $p=12,e=0.81$ (this corresponds to $pM=0.1183$ AU).  The other 
panels show 500 days until plunge (top right), 100 days to plunge (bottom 
left) and 1 day until plunge (bottom right).  The inspiral depicted here 
corresponds to the second-from-the-left black curve in the 
Fig.~\ref{fig:chi0contour}. The orbital configuration in the bottom-left panel is near a 1:2 $r$-$\varphi$ resonance which, in principle, could provide a substantial kick to the linear-momentum of the binary \cite{vandeMeent:2014raa}. We have not attempted to explore this effect in this work.
\label{fig:snapshots}}
\end{figure}

Using the interpolated hybrid self-force we computed a set of trajectories of 
extreme-mass-ratio-inspirals using the osculating element equations.  In 
Fig.~\ref{fig:snapshots} we show snapshots of a sample high-eccentricity 
inspiral, computed with $M=10^6 M_\odot$, $\epsilon = 10^{-5}$ and 
$(p_0,e_0)=(12,0.81)$.  For a sense of scale, at the initial configuration 
the inspiral's apastron is at $\sim0.623$ AU and its periastron is at 
$\sim0.0654$ AU; the entire inspiral occurs in the strong-field regime.  
In this example the EMRI takes 2115.45 days to evolve to plunge, during which 
it undergoes $\sim50132$ periastron passages.  Computing this particular 
inspiral took a few minutes on a standard 3GHz laptop.

\begin{figure}
\includegraphics[width=8.5cm]{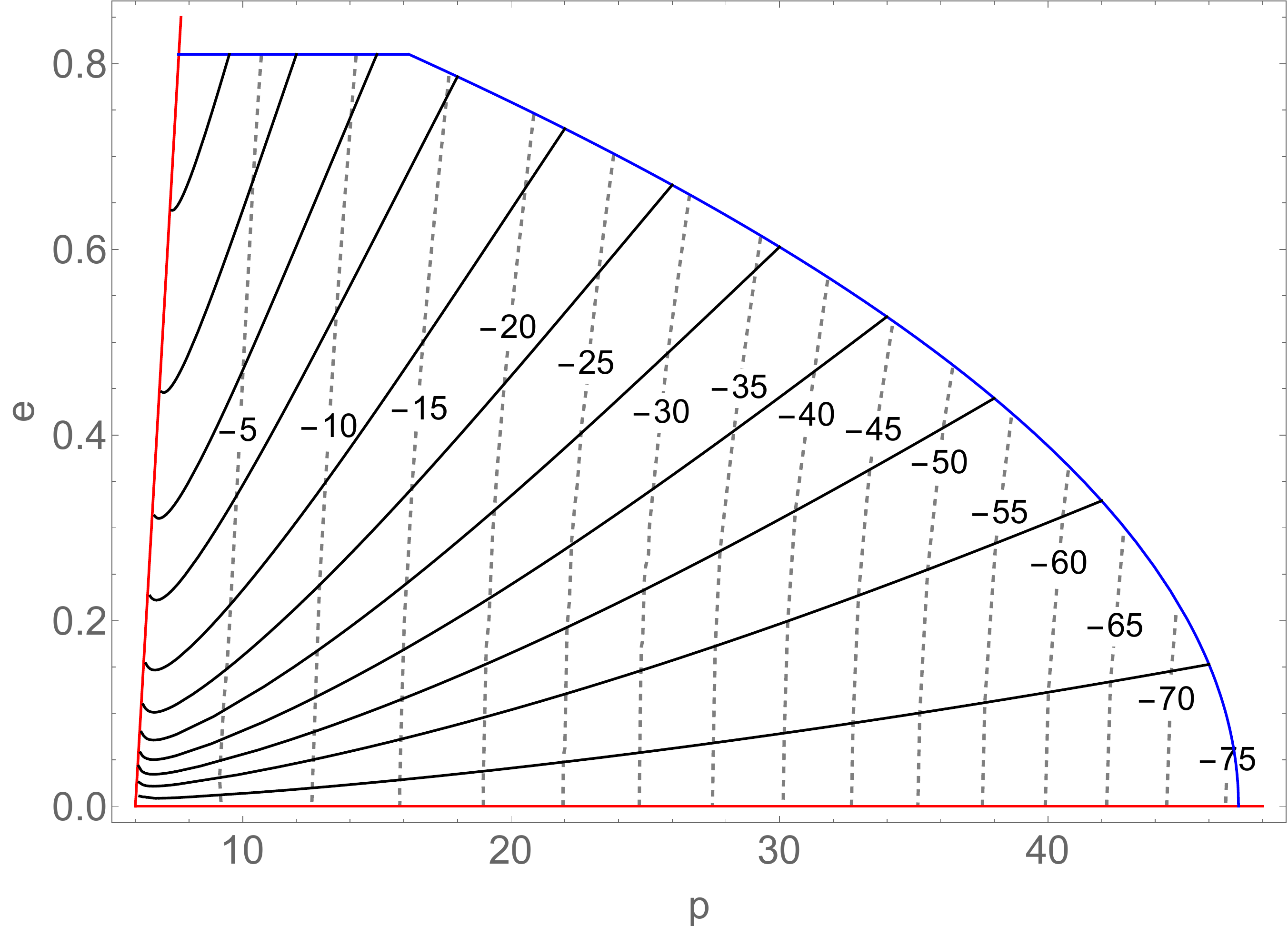}
\caption{Sample inspirals for $\mu/M=10^{-5}$ and $M=10^{6}M_\odot$.  Solid black 
curves show the evolution of $(p,e)$ from entering the LISA-like passband 
(marked with the blue curve). We truncate this curve to a constant in $e$ for $p\lesssim16$
as it is predicted that the initial eccentricity of EMRIs will not be above $\sim0.81$ \cite{HopmAlex05}.
Generally, as each inspiral progresses, both 
$p$ and $e$ decrease (with the exception of an increase in $e$ near the 
separatrix \cite{CutlKennPois94}).  The dashed lines are contours that mark 
the number of radians $\chi_0$ will evolve from a given point until plunge 
(this number is negative as the conservative self-force, and hence evolution 
of $\chi_0$, acts against the usual periastron advance \cite{BaraSago11}).  
\label{fig:chi0contour}}
\end{figure}

Over an inspiral the values of $p$ and $e$ generally decrease (with the 
possible exception of a small increase in eccentricity close to plunge 
\cite{CutlKennPois94}).  This behavior is best seen in a $(p,e)$ plot of 
the inspirals such as the one we show in Fig.~\ref{fig:chi0contour}.  In this 
figure we show the tracks in $p,e$-space of a number of inspirals from the 
point when they enter the observable band of a LISA-like spacecraft until 
plunge.  In addition we show, overlaid as a contour plot, the evolution of 
$\chi_0$.  As $\chi_0$ is predominantly affected by the conservative 
self-force, we can use $\chi_0$ to gauge the influence of the conservative 
self-force on an inspiral's phase.  We see that the conservative self-force 
subtracts somewhere between $10$ and $70$ radians of phase for an inspiral that 
starts with $p>14$.  Note that although the tracks in 
Fig.~\ref{fig:chi0contour} look very smooth, each track has many thousands 
of oscillations on the orbital time scale that are too small to appear on 
the plot.

The time to compute the tracks shown in Fig.~\ref{fig:chi0contour} varies 
greatly, with the shortest being a few minutes and the longest being tens of 
hours on a standard 3GHz laptop.  The reason for this large variation in 
computation time is that the self-force for orbits with a large value of 
$p$ is much smaller (e.g., for circular orbits $F^t$ scales as $r_0^{-5}$ 
for large $r_0$ \cite{BaraSago07}).  Consequently, inspirals evolve much more 
slowly in this regime.  For example, the bottom most track in 
Fig.~\ref{fig:chi0contour} starts with parameters $(p_0,e_0) = (46,0.152822)$ 
and goes through over $6\times10^{6}$ periastron passages before plunge.  
In contrast, the left-most track only goes through $\sim10^{3}$ periastron 
passages before plunge.  The latter takes minutes to compute whereas the 
former takes many hours.

In addition to computing inspirals for EMRIs we can also consider results 
for IMRIs.  For our evolution scheme to be valid the inspiral must evolve 
adiabatically, which will not be the case when $\epsilon$ is large and the 
particle is in the strong-field.  Reference~\cite{CutlKennPois94} quantified the 
allowed range of mass-ratios and found that so long as 
$\epsilon \ll (p-6-2e)^2$ the inspiral will evolve 
adiabatically\footnote{Near the ISCO this condition is modified to 
$\epsilon\ll (p-6)^{5/2}$}.  We thus see that even quite close to the 
separatrix our inspirals should be valid.  In addition, recent work has shown 
that the domain of validity of the conservative sector of perturbation 
theory likely includes IMRIs \cite{LetiETC11}.  For this reason we include 
an example IMRI inspiral.  As IMRIs evolve much faster than EMRIs, this gives 
us an opportunity to showcase the effects of the self-force on the inspiral 
on the orbital time scale.  We also take the opportunity to compare our 
inspiral computed using the full self-force with that computed using an 
adiabatic approximation and a dissipative-only approximation (matching the 
initial frequencies as outlined in Sec.~\ref{sec:freq_match}). 

\begin{figure}
\includegraphics[width=8.7cm]{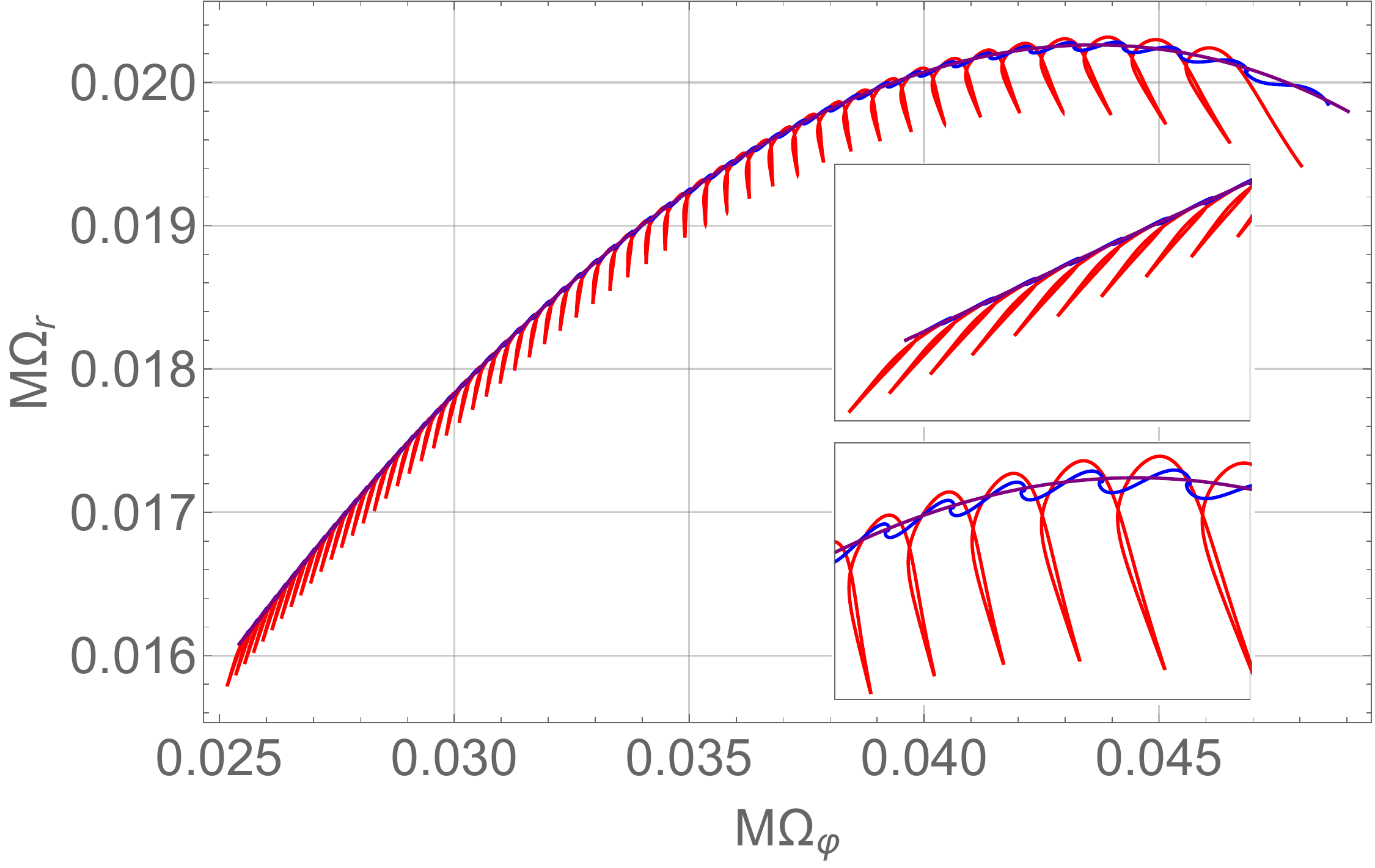}
\caption{Sample IMRI evolution.  The evolution of the orbital frequencies for 
inspirals computed using the full self-force $F^\a$ (red curve), the 
dissipative self-force $F^\a_\text{diss}$ (blue curve), and the adiabatic 
self-force $F^\a_\text{ad}$ (nonoscillatory, purple curve).  The mass ratio 
is $\e=5\times10^{-3}$.  The full self-force evolution is started with 
$(p,e)=(10,0.4)$ and the initial frequencies of the other inspirals are 
matched using the procedure outlined in Sec.~\ref{sec:freq_match}.  The upper 
inset shows the initial period of the evolutions where the different inspirals 
are in phase. The lower inset shows the inspirals close to plunge, where now 
the full and dissipative inspirals are starting to dephase.  At plunge the 
total discrepancy of the accumulated phase between the full and dissipative 
only inspirals is $\sim 2$ radians.
\label{fig:IMRI}}
\end{figure}

Our main result on IMRIs is presented in Fig.~\ref{fig:IMRI} where we show 
the evolution of the orbital frequencies for inspirals 
computed with $\epsilon=5\times10^{-3}$ and initial conditions $p_0=10$, 
$e_0=0.4$.  When the initial frequencies are matched, the full self-force 
inspiral, dissipative-only inspiral, and adiabatic inspiral initially evolve 
together.  The conservative self-force induces large oscillations in the 
orbital frequencies on the orbital timescale, whereas the dissipative only 
inspiral has smaller oscillations and the adiabatic inspiral exhibits no 
oscillations.  Even in this short inspiral, which lasts just 8 minutes, 
excluding the conservative self-force causes the inspiral to dephase by 2 
radians.

\section{Concluding remarks}
\label{sec:con}

In this paper we have computed high-eccentricity inspirals of a stellar-mass 
compact object into a massive Schwarzschild black hole while including all 
first-order-in-the-mass-ratio effects.  A key feature of this work over 
previous efforts is that we are able to model inspirals with an initially 
high eccentricity as they enter the detection band of a LISA-like spacecraft 
(previous work concentrated on the low-eccentricity case 
\cite{WarbETC12,Lackeos:2012de}).  This is important because it is expected 
that most observed EMRIs will initially be highly eccentric \cite{HopmAlex05}.

In computing inspirals we make use of a new code to compute the local 
Lorenz-gauge self-force acting on the particle \cite{OsbuETC14}.  Although 
this code is a marked improvement on previous codes 
\cite{BaraSago10,AkcaWarbBara13} in terms of speed and accuracy, the raw 
output of the code is not sufficiently accurate across the whole parameter 
space of inspirals to allow for the computation of inspirals with a phase 
error of less than $0.1$ radians.  To overcome this we note that the 
leading-order phase evolution is driven by the orbit-averaged fluxes radiated 
from the particle.  This inspires a hybrid scheme of combining the 
Lorenz-gauge results with fluxes calculated from a highly accurate RWZ code.  
The hybrid self-force is then precomputed to densely cover a wide region of 
orbital parameter space.  We are then able to implement a relatively local 
interpolation scheme for the self-force to rapidly compute extreme and 
intermediate-mass-ratio inspirals.  Typically an inspiral starting in the 
strong field will take a few minutes on a standard 3GHz laptop to evolve to 
plunge.  Our main results are presented in Sec.~\ref{sec:EMRI}.

Looking to the future, there are a number of open questions remaining.  
First, in order to complete the inspiral model, accurate to less than order 
unity in the phase evolution, it will be necessary to include 
second-order-in-the-mass-ratio effects \cite{HindFlan08} (see 
Table~\ref{table:required_accuraries}).  Currently there are no calculations 
of the second-order self-force, but the necessary formalism has been laid 
\cite{Poun12a} and computational techniques are emerging 
\cite{WarbWard14,WardWarb15}.  Once the second-order orbit averaged 
dissipative self-force can be computed, the results are easily added to our 
self-force interpolation scheme and inspiral model.

It will also be important to quantify the effects of the ``geodesic 
self-force approximation.''  The true self-force is a functional of the 
entire past history of the particle's motion but in our work we take the 
self-force at each instance to be that of a particle whose past history is 
motion along the tangent geodesic to the inspiralling worldline.  This 
approximation introduces a small error which is important to quantify.  
Initial investigations made by comparing a self-consistent evolution with a 
geodesic self-force evolution in the scalar case suggest this error is very 
small (with the phase error smaller than the error bars from either 
evolution \cite{Warb13,Warb14a,Dien15}).  Once self-consistent evolutions can be made in the gravitational 
case the results of our work here can be used for comparison to quantify 
the error from the geodesic self-force approximation.

In our work we concentrated on inspirals into a Schwarzschild black hole but 
it is expected that astrophysical black holes will generally be rotating.  
Thus it is important to extend inspiral models to motion around a Kerr black 
hole.  There has been much progress recently on computing self-forces in Kerr 
spacetime \cite{KeidETC10,PounMerlBara13,Merlin:2014qda,VandShah15} and these 
results can be used to compute inspirals in much the same way as we have done 
here.  Orbits in Kerr spacetime are generally computed in a radiation gauge. 
Thus, even in Schwarzschild spacetime, it would be interesting to compare 
an evolution computed using a radiation gauge self-force with our evolution 
computed using a Lorenz-gauge self-force.  Whilst the coordinate descriptions 
of the two evolutions might differ, the phase evolution should be the same.

Finally, we note that although we can rapidly compute inspirals in a matter 
of minutes, this is probably still not quick enough for use in practical 
matched filtering searches.  A similar problem is encountered when evaluating 
the time-domain EOB models for use in gravitational-wave searches with LIGO 
data.  One successful technique that has been applied in that case is the 
use of reduced order modelling \cite{Field:2013cfa} that allows for 
interpolation and rapid evaluation of the EOB waveforms.  No doubt a similar 
approach would be beneficial for more extreme mass ratios as well.

\acknowledgments

We thank Chad Galley, Scott Field, Barry Wardell and Seth Hopper for helpful 
discussions. We also thank Sarp Akcay for a careful reading and comments on a draft of this paper.
This work was supported in part by NSF Grant No. PHY-1506182.  
N.W.~gratefully acknowledges support from a Marie Curie International Outgoing 
Fellowship (PIOF-GA-2012-627781). T.O.~gratefully acknowledges support from 
the North Carolina Space Grant's Graduate Research Assistantship Program and 
a Dissertation Completion Fellowship from the UNC Graduate School.  
C.R.E.~acknowledges support from the Bahnson Fund at the University of North 
Carolina-Chapel Hill.

\bibliography{inspiral}

\end{document}